\documentclass[pre,twocolumn,noshowpacs]{revtex4}
\usepackage{epsfig}
\usepackage[]{natbib}
\usepackage{xspace}
\usepackage{color}
\usepackage{bbold}
\usepackage{tabu}
\usepackage{mathtools}

\begin{document}

\title{Ergodic properties of Brownian motion under stochastic resetting
}

\author{E. Barkai$^{1}$, R. Flaquer-Galm\'es$^{2}$,  V. M\'endez$^{2}$}
\affiliation{$^{1}$Department of Physics, Institute of Nanotechnology and Advanced Materials, Bar Ilan University, Ramat-Gan
52900, Israel\\
$^{2}$Grup de F\'{\i}sica Estad\'{\i}stica, Departament de F\'{\i}sica. Facultat de Ci\`{e}ncies, Universitat Aut\`{o}noma de Barcelona, 08193 Barcelona, Spain.}

\begin{abstract}

We study ergodic properties of one-dimensional Brownian motion  with
resetting. Using  generic classes of statistics
of times between resets, we find respectively for  thin/fat tailed distributions,  the normalized/non-normalised invariant density of this process. 
The former case corresponds to known results in the resetting literature
 and the latter to infinite ergodic theory. 
Two types of ergodic transitions are found in this system. The first is when
the mean waiting time between resets diverges, when standard ergodic theory switches to infinite ergodic theory.  The second is when the mean of the square root of time between resets diverges and the properties of the invariant density 
are drastically modified. We then find a fractional integral equation describing the density of particles. This finite time tool is particularly useful close
to the ergodic transition where convergence to asymptotic limits is logarithmically slow.  Our study implies rich ergodic behaviors for this non-equilibrium process which
should hold far beyond the case of Brownian motion analyzed here.

\end{abstract}

\maketitle

\section{Introduction}

 Stochastic processes under sporadic resetting gained considerable attention \cite{Evans2020,Gupta2022}.
Under certain conditions a non-equilibrium
stationary state (NESS) is found while the system still has non zero currents \cite{Evans2011,Pal2016,Fuchs2016,Eule2016}. NESS was
studied extensively for many processes with resetting
\cite{Evans2020,Gupta2022}, for example
for  Brownian motion (BM)  \cite{Evans2011,Friedman2020} and run and tumble processes
\cite{Martin2018}.
   In this work we investigate the ergodic properties of such a process
\cite{WeiWang2021,WeiWang2022,Stoj2022}. At the first stage
of our work we 
discuss a connection between the theory of  NESS and statistics of
 renewals, in particular  we study a useful relation between the resetting
 problem and the so called backward recurrence time \cite{Godreche2001}.
This not only gives a simple point of view on the  emerging NESSs, but can be
used to relate this timely problem to Dynkin's  backward time 
limit theorem \cite{Dynkin1955}  and its extension \cite{Wanli2018}. 

We study
BM, with times between resetting being independent identically
distributed (IID)  random variables (RVs)
  \cite{Pal2016,Eule2016,Gupta2016,Radice2022}.
 When the process, is thin/fat tailed respectively, we find the normalized/non-normalized invariant density of this system. 
Using Laplace transforms, Pal, Kundu and Evans \cite{Pal2016} and independently
Eule and Metzger \cite{Eule2016}  found
the normalized NESS of this process. 
Our work sheds light on these normalized
states, by connecting them to mathematical limit theorems from the field
of 
 renewal theory, but our main contribution is with respect to the less
understood non-normalized phase.  

Non-normalized states, were previously studied in the context of infinite ergodic
theory, both in the math \cite{aaronson1997}
and the physics literature \cite{akimoto2013,Aghion2019,Akimoto2020,Barkai2021,Giordano,Afek}.
 Here, our goal is to 
show how  this tool is used in the context of the resetting paradigm. 
More specifically we
find novel  ergodic transitions in this system. The first is anticipated,
and it is found  when the mean time between resetting diverges. The second,
takes place when the mean of the square root of time between resetting diverges.
In this case the properties of the infinite measure are modified, and
so are the relations between time and ensemble averages. 
As explained below, this second  transition is related to a competition between two mechanisms
of return to the origin, namely will the resetting control the return process,
or will it be the diffusion process itself? 
So our goal is to explain the rich phase diagram of the
ergodic properties in this system. 
 We focus
on the most well studied case, the underlying motion being BM, 
still while limiting our-selves to an example, the tools presented are
general. 

\section{Model and formal solution}

We start with a simple relation between the density of the  reset free
process and the density of the process with resetting. 
For that aim we will use three probability density functions (PDFs). 
Let
 $f(B,t)$ be the PDF of the backward recurrence time $B$ at time $t$, 
$\rho(x,t)$ the PDF
of the position $x$ of the particle at time $t$, and $G(x,t)$ the Green function
of the walker in the absence of reset. We now explain the basic properties of
these functions and their significance.  

A tagged particle
performs  one-dimensional
 BM
between resetting events,  hence
\begin{equation}
G(x,t)= { \exp\left(  - x^2 / 4 D t \right) \over \sqrt{ 4 \pi D t}}
\label{eq01}
\end{equation}
and $D$ is the diffusion constant. 
This is the propagator of a free BM without resetting, the particle starting
on the  origin $x=0$, at $t=0$.
The resetting is to the position $x=0$ which is also the origin of the
process.
The waiting times between resetting events are independent identically
distributed  (IID) random variables (RVs) drawn from a common
PDF of the waiting times $\psi(\tau)$. Thus at time $t=0$ the particle starts on the origin
$x(t)|_{t=0}= 0$, we draw a positive resetting time from $\psi(\tau)$ denoted $\tau_1$, 
the particle
performs a free BM  in the interval of time $(0, \tau_1)$, finally reaching some random
position $x(\tau_1 ^{-})$ (the superscript $-$ indicates a time just prior to the reset). Then the particle's position  is reset to zero
 $x(\tau_1 ^{+})=0$,
the process is then renewed, namely we draw a second waiting time $\tau_2$ also
from $\psi(\tau)$ etc. When this process is continued we
get the sequence of IID RVs $\{\tau_1,\tau_2,\tau_3 \cdots \}$,
i.e. the waiting times between resetting events, which are needed to construct
the path of the particle.  

%%%%%%%%%%
\begin{figure}[htbp]
	\includegraphics[width=1\hsize]{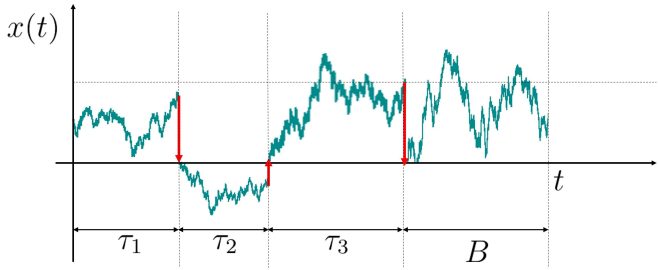}
	\caption{Schematics of BM with resetting. Since BM is Markovian the backward recurrence time $B$
controls the position of the restarted walker at time $t$. For short times,
statistical properties of $B$ will obviously depend on $t$, the only exception is when the waiting times are exponentially distributed. If the time $t$
 is long
a steady state of the random variable $B$ is found provided that
 the mean time between resets is finite.
This in turn will determine the steady state
 statistical properties of the stochastic
process $x(t)$. When the mean time between resetting events
diverges we are in the domain of infinite ergodic theory. }
	\label{fig1}
\end{figure}
%%%%%%%%%%%%%

 We are interested in the PDF of the position of the particle at time 
$t$
denoted $\rho(x,t)$. Let $x(t)$ be the stochastic process describing the location of the particle.   Since  BM is a Markovian process, hence $x(t)$
is connected to the time the last reset to $x=0$ was made.
This last reset event is at time $t-B$ and $B$ is called
the backward recurrence time (see schematics in Fig. \ref{fig1}). 
Clearly $B$ is a RV, whose statistical properties in general depend on $t$
while $0\le B \le t$.
 In the process just described $x(t)= x^f(B)$ where
$x^f(B)$ is the position of a reset free BM at time $B$, with the initial
position $x=0$. 
Hence from the well known properties of BM,  $x(t)$ for the resetting
process,  is a product of two independent random variables $\sqrt{B}$ and $\xi$,
namely
\begin{equation}
 x(t)= (D B)^{1/2}\xi 
\label{eqADD1}
\end{equation}
where $\xi$ is a Gaussian 
RV with zero mean and variance $\langle \xi ^2\rangle=2$. 

 The backward time is defined according to 
\begin{equation}
\sum_{i=1} ^N \tau_i + B =t,
\label{eqBack}
\end{equation}
where $N$ is the random number of resets in the time interval $(0,t)$,
see again schematics in Fig. \ref{fig1}.
Then using the  Fourier $x \to k$  transform of $\rho(x,t)$ 
and 
Eq. (\ref{eqADD1}),
$ {\cal F}[ \rho(x,t)]=\langle \langle \exp( i k \sqrt{D B} \xi)\rangle_{\xi}
\rangle_{B}= \langle \exp( - k^2 D B)\rangle_B$ were we used
the fact that $B$ and $\xi$ are independent RVs, and that
the PDF of $\xi$ is Gaussian. We then have 
\begin{equation}
 {\cal F}[ \rho(x,t)] = \int_0 ^{t} f(B,t) \exp( - k^2 D B) {\rm d} B
\end{equation}
and as mentioned $f(B,t)$ is the PDF of $B$. 
Hence inverting back to $x$ space,
the formal solution to the problem reads
\begin{equation}
\rho(x,t) = \int_0 ^t f(B,t) G(x,B) {\rm d} B.
\label{eq03}
\end{equation}
Luckily statistics of $B$ are well studied in the context of renewal theory, in particular 
the PDF $f(B,t)$ is studied in \cite{Godreche2001}.
Specifically, the Laplace transform of $f(B,t)$ is given in terms of the Laplace
transform of $\psi(\tau)$ in \cite{Godreche2001} and some further details will be provided below.

 To study non-equilibrium steady states we soon focus
on the long time limit of Eq. (\ref{eq03}). 
Before doing so we note that the approach is not limited to BM
in dimension-one, in essence many other 
transformation of $B$ might be considered, for example
consider BM in a force field \cite{Pal2015,Ray2020} with resetting, or  anomalous diffusion \cite{Masolivier2019,Bordova2019,Bordova2020,Mendez2021,Majumdar2022,Weron2022},
 or deterministic processes etc.  
And as explained below the waiting time strategy just described, is identical
to a much studied  time varying rate approach \cite{Pal2016,Eule2016}. 
In all these problems the backward recurrence time plays a crucial role
hence we will soon recap some of its properties. 

 In this paper we will focus on two classes of resetting processes.
The first are processes with a  smooth PDF $\psi(\tau)$ of the reset times,
when the positive  integer moments of the waiting times  are finite.
The most  studied example is $\psi(\tau) = \exp(- \tau)$ for $\tau>0$
and we have set here the mean waiting time to be unity. We will
also discuss briefly deterministic 
resetting $\psi_{{\rm det}} (\tau)=\delta(\tau- \tau_{{\rm det}})$, which is 
a special case.  
The second class of PDFs have a power law tail
for large $\tau$
\begin{equation}
\psi(\tau) \sim (\tau_0)^\alpha \tau^{- 1 -\alpha}, \ \ \mbox{and} \ \ 0<\alpha<1.
\label{eq04}
\end{equation}
As well known these PDFs belong to the domain of attraction of L\'evy's laws and
 the big jump principle \cite{Burioni}
 holds instead of standard large deviation theory. 
In particular the mean waiting time diverges and in that sense the process is scale free. Other cases, like $\alpha=1$ and $1<\alpha<2$
are also of interest, but due to space considerations will not be presented here.

\section{Statistics of the backward recurrence time}

 In the long time limit the measurement time $t$ typically falls in
a time interval which is longer than the average, see viewpoint on
this issue in \cite{Pal2022}. This time interval straddling
time $t$ is a sum of the forward recurrence time (time between $t$ and
the first renewal after $t$) and the mentioned backward time (time
between $t$ and the previous reset event). The steady state (ss)
PDF of $B$ for non-lattice PDFs of resetting time
is \cite{Godreche2001,Dynkin1955} 
\begin{equation}
f^{{\rm ss}} (B) = { 1 - \int_0 ^B \psi(\tau) {\rm d}  \tau \over \langle \tau \rangle}
\label{eq05}
\end{equation}
and $\langle \tau \rangle = \int_0 ^\infty \tau \psi(\tau) {\rm d} \tau$ is the mean time between resetting. 
When $\psi(\tau)$ is exponentially decaying,  the 
distribution of 
 $B$ is $f(B)= \exp( - B)$, and hence  the same as that of the 
distribution of the time between
resetting. 
 More precisely in the long time limit $f(B,t)$ reaches
this steady state, provided that the mean time between resets $\langle \tau \rangle$ is finite (class number one above)  and hence
\begin{equation}
\lim_{t \to \infty} \langle \tau \rangle f(B,t) = 1 - \int_0 ^B \psi(\tau) {\rm d} \tau= {\cal I}^{{\rm ss}} (B).
\label{eq06}
\end{equation}
Here we call ${\cal I}^{{\rm ss}}(B)$ the invariant steady state which is dimensionless and also a perfectly normalizable function. What happens when $\langle \tau \rangle$ diverges?

 If $\alpha<1$ (class 2) then as shown by Wang et al.  the PDF of $B$ satisfies \cite{Wanli2018}
\begin{equation} 
\lim_{t \to \infty} \langle \tau^*(t) \rangle f(B,t) = 1 - \int_{0} ^{B} \psi(\tau) {\rm d} \tau= {\cal I}^{\infty}(B).
\label{eq07}
\end{equation}
For a brief recap of this and other basic results see Sec. \ref{SecUni}.
Eq. (\ref{eq06}) and Eq. (\ref{eq07}) appear similar, but they are not. In Eq. (\ref{eq07})
 $\langle \tau^*(t) \rangle\propto t^{1-\alpha}$ is increasing with measurement time (see below). Still
the invariant densities ${\cal I}^{{\rm ss}}(B)$ and ${\cal I}^{\infty}(B)$ have the
same functional dependence on the waiting time PDF,  though
${\cal I}^\infty(B)$ is called an infinite invariant density since it is not a normalizable
function. Since by definition on the left hand side of Eq. 
(\ref{eq07})  we take a perfectly normalised  function
$f(B,t)$ and multiply it  by a monotonically increasing function
of time and take the long time limit, the integration of 
${\cal I}^{\infty}(B)$ over $B>0$ diverges. This is because ${\cal I}^{\infty}(B)\propto B^{-\alpha}$ and $0<\alpha<1$ and hence it is a non-integrable function due 
to its large $B$ behavior. 

Eqs. (\ref{eq06})  and (\ref{eq07}) are valid for any finite $B$ in the limit of long measurement
times. However, especially when $0<\alpha<1$ 
the case when $B$ scales with measurement time
must also be considered, namely when $B \propto t$ and $t$ is made large.
This limit was studied by Dynkin who found \cite{Dynkin1955,Godreche2001}
\begin{equation}
f(B,t) \sim {1 \over t} {\rm Dyn}  \left( { B \over t} \right),
\label{eq08}
\end{equation}
and the scaling function reads
\begin{equation}
{\rm Dyn} (y)= { \sin \pi \alpha \over \pi} { 1 \over y^\alpha (1- y)^{1-\alpha}},
  \  \  \  0<y<1. 
\label{eq09}
\end{equation}
This formula shows that the most likely events, are obtained when
$y\sim 0$ or $y \sim 1$, where the Dynkin PDF diverges, corresponding
to either very short $B$ compared to $t$ or $B$ of the  order of $t$. 
Note that when $\alpha=1/2$ we find the arcsine law attributed to P. L\'evy.

 We see that for $\alpha<1$ we have two limiting laws, one for
$B$ fixed and measurement time long, and the other when the ratio $B/t$ is
fixed. The use of these laws, for example for the calculations of expectation
values, depends on the observable of interest. We will later study observables
which are integrable with respect to the infinite density, and 
show how infinite  ergodic theory plays a special role for the non-equilibrium
steady states.

Note that Eq. (\ref{eq07})
and Eq. (\ref{eq09})  are related as they have to match.
To see this, using Dynkin limit theorem, with $B \ll t$ we have
\begin{equation} 
f(B,t) \sim {1 \over t} {\sin \pi \alpha \over \pi} {1 \over (B/t)^\alpha}.
\end{equation}
On the other  hand using Eq. (\ref{eq04}),
${\cal I}^{\infty} (B) \sim (\tau_0)^\alpha B^{-\alpha}/\alpha$ for large $B$. Since $f(B,t) \sim {\cal I}^{\infty} (B)/\langle \tau^*(t) \rangle = (\tau_0)^\alpha B^{-\alpha} / \alpha \langle \tau^*(t) \rangle$ we easily find
\begin{equation}
\langle \tau^{*}(t) \rangle = { \pi (\tau_0)^\alpha \over \alpha \sin \pi \alpha} t^{1-\alpha}.
\label{eq12}
\end{equation}
This is exactly the expression found in \cite{Wanli2018}. 
Note that roughly speaking
$\langle \tau^*(t) \rangle$ is a mean time between resets, in the sense that
if we integrate only up to $t$  $\langle \tau^*(t) \rangle \propto \int^t \tau \psi(\tau) {\rm d} \tau \sim t^{1-\alpha}$ as indeed we have found. 
More precisely, let $\langle N \rangle$ be the averaged number of resets 
in the time interval $(0,t)$. Then in the long time limit
\cite{Godreche2001} 
\begin{equation}
{{\rm d} \langle N \rangle \over {\rm d} t}  \sim \left\{
\begin{array}{c c}
{1 \over  \langle \tau^{*}(t) \rangle} & {\mbox when} \ \  0<\alpha<1 \\    
{1 \over \langle \tau \rangle} & {\mbox otherwise}.
\end{array}
\right.
\label{eq12aaa}
\end{equation}
Thus $1/\langle \tau \rangle$ and $1/ \langle \tau^*(t) \rangle$ are the long time
rates of the underlying  renewal process. Namely $d \langle N \rangle $, which is the probability of observing a reset event in the time interval $(t,t+ {\rm d} t)$, is given by ${\rm d}t /\langle  \tau^*(t) \rangle$ for $0<\alpha<1$ and by ${\rm d} t /\langle \tau \rangle$  otherwise. 
Thus using  Eqs. (\ref{eq06},\ref{eq07},\ref{eq12}) we summarize
\begin{equation}
\lim_{t \to \infty} {f(B,t) \over  \left( {{\rm d} \langle N \rangle \over  {\rm d} t} \right)} =
S(B) 
\label{eq13}
\end{equation}
and  $S(B)$ is the survival probability, i.e. the probability of not performing a reset
in time $B$  
\begin{equation}
S(B) = 1 - \int_0 ^B \psi(\tau) {\rm d} \tau = \int_B ^\infty \psi(\tau) {\rm d} \tau.  
\label{eq14}
\end{equation}
Eq. (\ref{eq13})  gives the invariant density of the backward time, be it either normalizable
or not.

\section{NESS}
\subsection{Normalized invariant density}

We now consider thin tailed waiting time PDFs of the first class.
The non-equilibrium steady state, $\rho^{{\rm ss}}(x)$ is 
 based
on the long time limit of the distribution of $B$ using Eqs. (\ref{eq03},\ref{eq05},\ref{eq14})
\begin{equation}
\lim_{t \to \infty} \langle \tau \rangle \rho(x,t) = \langle \tau \rangle \rho^{{\rm ss}}(x) = \int_0 ^\infty S(B) { \exp\left( - {x^2 \over 4 D B}\right) \over \sqrt{ 4 \pi D B} } {\rm d} B.
\label{eq15}
\end{equation}
For example setting $D=1/2$  and using
 $\psi(\tau)= \exp(- \tau)$ so $\langle \tau \rangle=1$ and
hence $S(B)= \exp( - B)$ we find the result in \cite{Evans2011}
$\rho^{{\rm ss}}(x) = \exp[ - \sqrt{2} |x|]/\sqrt{2}$, which exhibits the
typical non-analytical behavior at $|x|\to 0$. The latter
is a rather general feature of NESS,  since if we expand the Gaussian
in Eq. (\ref{eq15})   
to second order in $x$, the $x^2$ term will diverge, since 
$S(0)=1$ and $B^{-3/2}$ is non integrable at $B \rightarrow 0$.
Pal et al, \cite{Pal2016} derived  a formula for the steady state, 
which is identical to
Eq. (\ref{eq15}) without invoking the backward
recurrence time and using Laplace transforms (see also \cite{Eule2016}).
In Appendix A we make the comparison between the two results and
explain the different notations. 
%The model used 
%in \cite{Pal2016}  considers a  resetting protocol with a time varying rate,
%where the resetting rate $r(t)$ is a function of time since the last reset
%event. The rate  model in \cite{Pal2016} is identical to our approach (see Appendix A for more details). 
%

\subsection{Non-normalized invariant density $1/2<\alpha<1$}

 For the second class of PDFs given in Eq. (\ref{eq04}), we again start
with Eq. (\ref{eq03}) which gives $\rho(x,t) = \langle G(x, B) \rangle$,
here the average is with respect to the distribution of the backward time
$B$. As we soon explain, the  Gaussian function
 $G(x,B)$ can be either integrable with respect
to the infinite density of $B$ when $1/2<\alpha<1$ or not corresponding to  $0<\alpha<1/2$. The
case $\alpha=1/2$ marks a transition in the ergodic properties of the 
system. This is in addition to the marginal case $\alpha=1$
that marks a second transition to a normalized steady state
and perfectly standard ergodic theory. 

 Consider $1/2<\alpha<1$ and insert Eq. (\ref{eq07}) in Eq. (\ref{eq03})
using Eq. (\ref{eq14}).
We find an expression which looks similar to Eq. (\ref{eq15})
\begin{equation}
\lim_{t \to \infty} \langle \tau^{*}(t) \rangle  \rho(x,t) = \int_0 ^\infty S(B) { \exp\left( - {x^2 \over 4 D B } \right) \over \sqrt{ 4 \pi D B} } {\rm d}B = \tilde{{\cal I}}^{\infty} (x). 
\label{eq16}
\end{equation}
Of course the major difference if compared with Eq. (\ref{eq15}) is that
on the left hand side of Eq. (\ref{eq16})  we have the effective time dependent
mean resetting
time $\langle \tau^*(t) \rangle$. 
To see that the integral on the right hand side  exists, 
namely that $G(x,B)$ is indeed  integrable with respect
to the non-normalized state ${\cal I}^\infty(B)$ if $1/2<\alpha<1$, 
note that for large
$B$,
$S(B) \propto B^{- \alpha}$ 
and hence the integral in Eq. (\ref{eq16}) converges or
diverges if $1/2<\alpha<1$ or $0<\alpha<1/2$, respectively.
The function $\tilde{{\cal  I}}^{\infty} (x)$ defined in Eq. (\ref{eq16})
is  the non-normalizable non-equilibrium
 steady state
of $x$, as the integration over $x$ of this function diverges.  
The formula is valid for any finite $x$ in the long time limit. 
Note that we use the convention that the argument in the parenthesis defines
the infinite density of interest, thus ${\cal I}^\infty(B)$ is the infinite density of $B$ while $\tilde{{\cal I}}^{\infty}(x)$ is the infinite density of $x$. 

\subsection{Scaling solution $0<\alpha<1$}

 When $x \propto \sqrt{t}$  and both $x$ and $\sqrt{t}$ are large a different approach
is needed. Inserting  Dynkin's limit theorem Eq.  
(\ref{eq08}) in Eq. (\ref{eq03}) we find
\begin{equation}
\rho(x,t) \sim {1 \over t} \int_{0} ^t \mbox{Dyn} \left( { B \over t} \right) G(x,B) {\rm d} B.
\label{eq17}
\end{equation}
Making this equation explicit, we use Eqs. (\ref{eq01},\ref{eq09}) and simple
change of variables to obtain
\begin{equation}
\rho(x,t) \sim {g_\alpha \left(\xi \right)
\over \sqrt{ 2 D t} } \ \  \mbox{with} \ \ \xi = |x|/\sqrt{ 2 D t}.
\label{eq18}
\end{equation}
The scaling here is $x \propto \sqrt{t}$ hence it is diffusive, 
and this limit describes what we may call
the typical events, further it is valid for $0<\alpha<1$. 
 The scaling function is 
\begin{equation}
g_\alpha ( \xi) = { \sin \pi \alpha \over \pi}
 \int_{0} ^{1}  { \exp\left( - { \xi^2 \over 2 \eta} \right) \over \eta^\alpha (1 - \eta)^\alpha } { {\rm d} \eta \over \sqrt{ 2 \pi \eta} }.
\label{eq19}
\end{equation}
This  function unlike the invariant densities Eqs. 
(\ref{eq15},\ref{eq16}) does not depend on the fine details
of the model, i.e. on the waiting time PDF, beyond the parameter $\alpha$.
After change of variables, we find
\begin{equation}
g_\alpha(\xi) = { 1 \over \sqrt{ 2 \pi} \Gamma(1- \alpha)} 
U\left(\alpha, {1 \over 2} + \alpha, {\xi^2 \over  2} \right)e^{ - \xi^2/2}
\label{eq20}
\end{equation}
where we used the Tricomi function also called the Kummer
function of the second kind. This equation was derived with a different approach by Nagar and Gupta \cite{Gupta2016}. Here we have emphasized the connection between the resetting problem and  Dynkin's limit theorem.  
We think this is worth while, since in many fat tailed resetting problems,
the scaling solution for the diffusing particle
 will depend on this law, for example
if we replace the Gaussian propagator of free diffusion $G(x,B)$
 with a propagator of
anomalous type, similar laws will follow.

 The behavior of the scaling solution in vicinity of the resetting point
$x=0$
is of interest.
Exploiting the small $\xi$ limit of the Kummer function we have
\cite{Abr}
\begin{equation}
g_\alpha(\xi) \sim \left\{
\begin{array}{c c}
{2^{\alpha-1} \over \sqrt{ \pi}} { \Gamma\left(\alpha- {1 \over 2} \right) \over \Gamma\left( 1 - \alpha\right) \Gamma\left( \alpha \right) } {1 \over  \xi^{2 \alpha -1 }}  & \  {1 \over 2}  < \alpha<1 \\
\ & \ \\
- { 1 \over \sqrt{2} \pi^{3/2}} \left[ 2 \ln \xi - \gamma - 3 \ln 2\right] &
\ \alpha={1 \over 2} \\
\ & \ \\
{\Gamma\left( {1 \over 2}  - \alpha\right) \over \sqrt{2} \Gamma(1 - \alpha) \pi} & \ 0<\alpha< {1 \over 2} 
\end{array}
\right.
\label{eq21}
\end{equation}
where $\gamma$ is the Euler-Mascheroni constant.
We see that the scaling solution, when $\xi \to 0$,
 exhibits a transition at $\alpha=1/2$.
When $0<\alpha<1/2$ the scaling function at $x=0$, namely $g_\alpha(\xi=0)$ is a constant, and as
Eq. 
(\ref{eq20}) shows
this constant diverges when $\alpha\to 1/2$ from below. 
Further when $\alpha \to 0$ we find $\rho(x=0,t)\sim g_0 (\xi=0)/\sqrt{ 2 D t}= 1/ \sqrt{ 4 \pi D t}$ which is  the expected result since for $\alpha\to 0$ the solution $\rho(x,t)$ is 
the Gaussian PDF describing free BM.

 As a stand alone, Eqs. 
(\ref{eq18},\ref{eq21})
indicate that  $\rho(x,t) \to \infty$ when $x\to 0$ (or $\xi\to 0)$ and when $1/2<\alpha<1$. 
Clearly this is an unphysical effect. The density of the particles
$\rho(x,t)$, for thin tailed distributions on the origin $x=0$,
 is always finite for any $t>0$,
 and with power law distributed times between the resetting, we expect an even
lower density, since particles can escape to larger distances. 
 Mathematically, the small  $x$ regime is 
exactly the regime where the infinite density solution, namely Eq. (\ref{eq16}) with $\rho(x,t) \sim \tilde{{\cal I}}^{\infty}(x)/ \langle \tau^*(t) \rangle$  plays an important role
 (recall that the latter formula  is valid for finite $x$). In this sense the
non-normalized state cures an unphysical feature of the scaling solution
Eq. 
(\ref{eq20}).
To study this issue let 
\begin{equation}
\psi(\tau) = \alpha (t_0)^\alpha \tau^{- (1 + \alpha)} \ \mbox{for} \ \tau> t_0,
\label{eqppaa}
\end{equation}
otherwise it is zero, hence from Eq. (\ref{eq04})
$(\tau_0)^\alpha = \alpha (t_0)^\alpha$.   
Using the non-normalized state Eq. 
(\ref{eq16}),
for $x=0$
\begin{equation}
\rho(0,t) \sim { 2 \alpha \over 2 \alpha -1} { \sin \pi \alpha \over \pi^{3/2}  }  { ( \tau_0/\alpha^{1/\alpha})^{1/2 -\alpha } \over \sqrt{D} t^{1-\alpha}}. 
\label{eqAADD}
\end{equation}
Here we see that when $\alpha\to 1$ the time dependence vanishes as we approach
the standard behavior of thin tailed statistics, 
and when $\alpha\to 1/2$ the solution decays like one over square root of time, as expected from diffusion. 
Returning to Eq. (\ref{eq21}), this equation 
is valid for length scales $x> C(\alpha) \sqrt{ D \tau_0}$ where $C(\alpha)$ depends on the exponent $1/2<\alpha<1$ and in principle it can be estimated by
equating Eq.  (\ref{eqAADD}) and Eq. (\ref{eq21}).

More technically, one can verify that the large $x$ expansion of $\rho(x,t)$ in  Eq. (\ref{eq16}), is exactly the same as the small $x$ limit of the scaling solution employing
Eq. (\ref{eq21}). Thus the infinite density solution for large $x$ matches
the scaling solution for  small $x$, as it should. 

\subsection{Non-normalized invariant density $0<\alpha<1/2$}

 Finally, what is the infinite density for $0<\alpha<1/2$? Using Eq. (\ref{eq18})
\begin{equation}
\lim_{t \to \infty} \sqrt{ 2 D t} \rho(x,t) = g_\alpha(0) = \tilde{{\cal I}}^{\infty} (x). 
\label{eq22}
\end{equation}
and the constant $g_\alpha(0)$ is given in Eq. 
(\ref{eq21}). Here the infinite density is $x$ independent 
 so it is clearly a non-normalizable function, as the integration over $x$ from
minus infinity to infinity diverges. The infinite density here 
does not depend on the structure of the  waiting time PDF, and hence is very 
different if compared to the invariant densities found for thin tailed distribution, namely the normalized state, and that found in Eq. (\ref{eq16}).

Note that in Eq. (\ref{eq16}) we used the ever increasing
 time scale $\langle \tau^{*}(t) \rangle$ to define the non-normalized state,
while for the case case $0<\alpha<1/2$ we used the diverging length scale $\sqrt{ 2 D t}$. These infinite densities are certainly not probability densities and their use will be explained later, in fact the units of the infinite
density can be either the inverse of time or inverse of length,
depending on the value of $\alpha$.  In general infinite densities are  defined up to some
arbitrary constant (since these functions are not normalized we have some
freedom in the definition). This does not pose any problem, as long as one recalls  the basic definitions. For example to visualise the infinite density in simulations, we plot the density $\rho(x,t)$ times $\sqrt{ 2 D t}$ for finite $x$ and increase time, the solution in the long time limit will approach $\tilde{{\cal I}}^\infty(x)$ for $0<\alpha<1/2$. Or we plot $\langle \tau^*(t) \rangle \rho(x,t)$ for finite range of $x$, and  then as we increase measurement time the
solution will approach the asymptotic infinite density Eq. (\ref{eq16}). 
Of course as
$t$ is increased, most of the particles are actually diffusing far from the origin. Thus in practice if  $t$ is too long and the number of trajectories in simulation not large enough, it will be hard to visualise the infinite densities.   
To meet this sampling challenge we plot now the non-normalized states
for representative case. 

%%%%%%%%%%
\begin{figure}[htbp]
	\includegraphics[width=1\hsize]{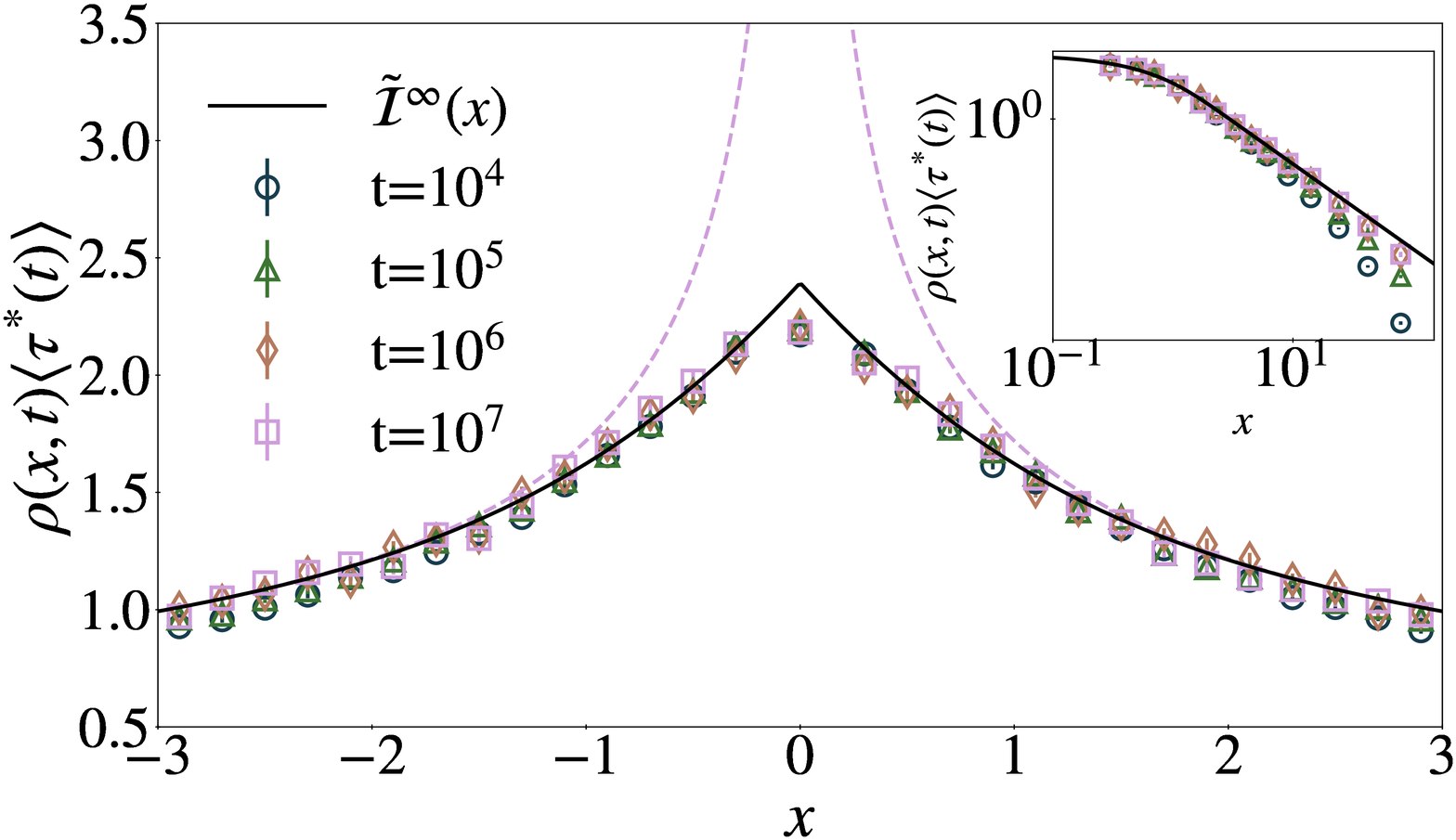}
	\caption{The infinite density versus $x$ for $\alpha=3/4$.
In simulation we plot the histogram for the
 density $\rho(x,t)$ multiplied by $\langle \tau^* (t) \rangle$ versus $x$.
Increasing time we see that results converge to  the theoretical prediction
namely the solid line presenting the infinite density $\tilde{{\cal I}}^\infty(x)$,
Eq.
(\ref{Infa304}). 
 For large $x$,  $\tilde{{\cal I}}^\infty(x) \simeq 1/|x|^{1/2}$, Eq. (\ref{eqlll}),
 hence this invariant density is not normalized. The infinite density exhibits a typical cusp on the origin.  Also shown is the scaling solution Eq. (\ref{eq21}) for time $10^7$
(dashed line).  The latter is a good approximation for diffusive scales, namely when  $x$ of the order $t^{1/2}$,
but for small $x$ presented here, the scaling solution clearly fails. 
 We use the Pareto PDF of times between
resets with $t_0=1$  and $D=1/2$. We have used $1.3\cdot10^6$ trajectories. \textit{Inset}: we present the large $x$, and the data for various times does not collapse on a master curve, unlike the small $x$ shown in the main figure. }
	\label{fig2}
\end{figure}
%%%%%%%%%%%%%

%%%%%%%%%%
\begin{figure}[htbp]
	\includegraphics[width=1\hsize]{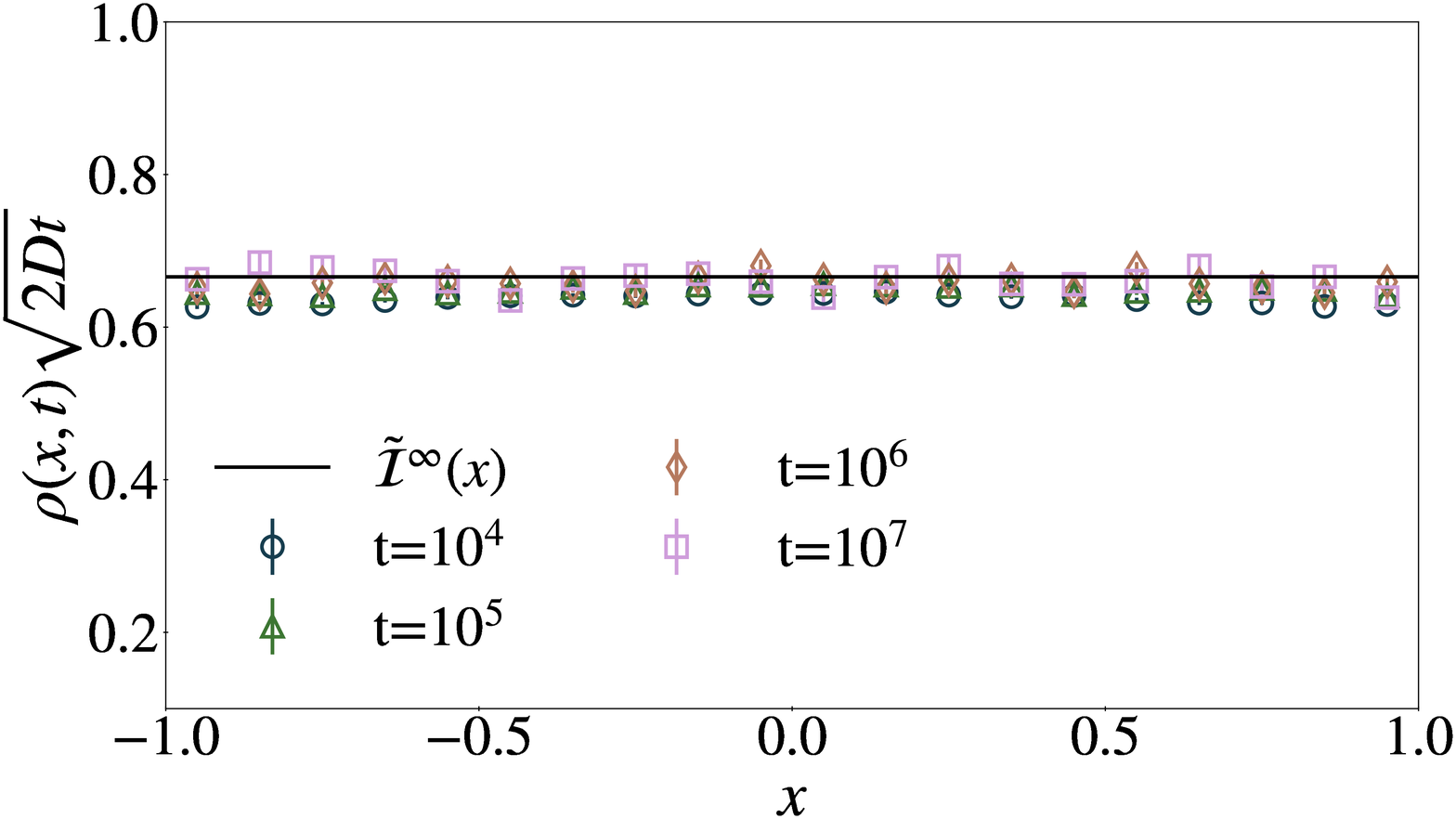}
	\caption{The infinite density versus $x$ for $\alpha=1/4$.
In simulation we plot the histogram for the
probability density $\rho(x,t)$ multiplied by $\sqrt{2 D t}$, unlike the case $\alpha>1/2$ where we use $\langle \tau^* (t) \rangle$, as shown in Fig. \ref{fig2}.
The solid line is the theory, the infinite density $\tilde{{\cal I}}^\infty(x)$
Eq.
(\ref{eq22}), namely $\tilde{{\cal I}}_{1/4} (x)= g_{1/4}(0)= \Gamma(1/4)/ \sqrt{2} \Gamma(3/4) \pi \simeq 0.6659$.
Now the infinite density is simply a constant, and the cusp found for $\alpha=3/4$ in Fig. \ref{fig2}, is not present.
 Further, clearly a constant invariant density is not normalized. 
 We use the Pareto PDF of times between
resets with $t_0=1$  and $D=1/2$. We have used $10^8$ trajectories.  }
	\label{fig3}
\end{figure}
%%%%%%%%%%%%%

\subsection{Examples}
\label{SecEx}

 As a specific example  consider the Pareto PDF of resetting times
$\psi(\tau)= \alpha \tau^{-1- \alpha} \  \mbox{when} \  1<\tau$
 otherwise 
$\psi(\tau)=0$, hence $t_0=1$ in Eq. 
(\ref{eqppaa}).
We also set $D=1/2$. In this case
\begin{equation}
\langle \tau^{*}(t) \rangle \sim (\pi / \sin \pi \alpha) t^{1-\alpha},
\label{eqttaauu}
\end{equation}
where we used Eqs. 
(\ref{eq04},
\ref{eq12}). It follows from
Eq. 
(\ref{eq16})
\begin{equation}
\tilde{{\cal I}}^{\infty} (x) =
\int_{0} ^{1}  { \exp\left( - {x^2 \over 2  B } \right) \over \sqrt{ 2 \pi  B} } {\rm d}B +
\int_{1}  ^{\infty} B^{-\alpha}   { \exp\left( - {x^2 \over 2  B } \right) \over \sqrt{ 2 \pi  B} } {\rm d}B. 
\label{eqINFpar}
\end{equation}
For $x=0$ we find
\begin{equation} 
\tilde{{\cal I}}^{\infty} (0) = \sqrt{ {2 \over \pi} } { \alpha \over \alpha - {1 \over 2} } . 
\end{equation}
Hence for $1/2<\alpha<1$ the density on the origin
\begin{equation}
\rho(0,t) \sim 
{ \tilde{{\cal I}}^{\infty} (0) \over \langle \tau^{*}(t)\rangle}
\end{equation}
is decreasing with time, and the divergence of the scaling solution 
Eq. (\ref{eq21}) at $\xi=x=0$
is not relevant, since as mentioned  that solution is not valid in this regime.
For the example  $\alpha=3/4$ 
Eq. (\ref{eqINFpar}) gives

\begin{eqnarray}
\tilde{{\cal I}}^{\infty}_{\alpha=3/4}  (x) &=&  
|x| \left( {\rm erf} \left( { |x| \over \sqrt{2} } \right) - 1 \right) + 
\sqrt{{ 2 \over \pi}} e^{ - {x^2 \over 2} }  \nonumber\\
&+&
{ \Gamma\left( {1 \over 4} \right) - \Gamma\left( {1 \over 4} , { x^2 \over 2} \right) \over 
2^{1/4} \sqrt{ \pi} \sqrt{ |x| } }
\label{Infa304}
\end{eqnarray}
where we use the error function and the incomplete Gamma function.
This invariant density is plotted in Fig. \ref{fig2} together with finite
time numerical simulations.  As shown the infinite density  exhibits a  cusp
on  the reseting point $x=0$, which is found also
 for resetting problems with a normalized
invariant density. 
For large $x$ we have
\begin{equation}
\tilde{{\cal I}}^{\infty}_{\alpha=3/4}  (x) \sim
{ \Gamma \left( {1 \over 4} \right) \over 2^{1/4} \sqrt{ \pi} \sqrt{ |x|}}.  
\label{eqlll}
\end{equation}
This expression can be shown to match the small $x$ behavior of the scaling solution Eq. (\ref{eq21}). Similarly in Fig. \ref{fig3} we study the case $\alpha=1/4$. As predicted by theory the infinite density is structureless, namely it is equal to a constant. This is clearly unlike what we see in Fig. \ref{fig2}, so the transition at $\alpha=1/2$ is evident.

\subsection{Efficient sampling}

When simulating invariant densities of restart processes one may follow at least two strategies.
The first is to construct the renewal process, namely using a simple program and a given $\psi(\tau)$,
we find the sequence $\{\tau_1, \tau_2, ..., B\}$, for a given measurement time $t$. Using $B$ it is
then easy to find $x_{{\rm free}}(B)$ which denotes the reset free dynamics at time $B$. As mentioned this is the same
as the coordinate of the walker with resetting at time $t$. This method is by far more efficient
if compared with continuously recording full trajectories of the reset process. This is clearly the case as
$B<t$. Further, if we have the measurement free propagator as for the case considered here,
then given $B$ one may find $x_{{\rm free}}(B)=x(t)$ without simulating the trajectory.
We have generated Fig. \ref{fig2} using both methods, and as expected the results are the same,
though the first method is by far more efficient.

\section{The moments}

\subsection{Thin tailed distributions}

 Using Eq. (\ref{eqADD1}) the moments of the process satisfy
\begin{equation}
\langle x^{2 m} (t) \rangle = D^m \langle \xi^{ 2 m } \rangle\langle B^m\rangle
\label{eqM01}
\end{equation}
and here $m=0,1,2, ..$. We used the fact that odd moments of $x(t)$ vanish
from symmetry. Recall that the PDF of $\xi$
 is Gaussian with zero mean and variance equal $2$ namely
$\mbox{PDF}(\xi) = \exp(- \xi^2/4)/\sqrt{ 4 \pi}$.
In the normalized steady state, namely when we deal with thin tailed PDFs of resetting
times, the moments of $B$ become time independent and so do the moments
of $x(t)$. For example 
\begin{equation}
\langle x^2 \rangle_{{\rm ss}}= 2 D \langle B \rangle_{{\rm ss}}.
\label{eqM02}
\end{equation}
and in general $\langle x^{2 m} \rangle_{\rm ss} = D^m \langle \xi^{2 m} \rangle \langle B^m \rangle_{{\rm ss}}$, where
\begin{equation}
 \langle  \xi^{2 m} \rangle= {4^m \Gamma(m+1/2) \over \sqrt{ \pi}}.
\label{eqM02N}
\end{equation}
 In turn the moments of $B$ are determined by the moments
of $\psi(\tau)$ using 
Eq.
(\ref{eq05}).
We use the Laplace transform  $\hat{\psi}(s) = \int_0 ^{\infty} \exp(- s \tau) \psi(\tau) {\rm d} \tau$, and similarly $\hat{f}_{{\rm ssB}}(s)$ is the Laplace pair
of $f_{{\rm ss}}(B)$. Using the convolution theorem and Eq. (\ref{eq05})
\begin{equation}
\hat{f}_{{\rm ssB }} (s) = { 1 - \hat{\psi}(s) \over s\langle \tau \rangle}.   
\label{eqM03}
\end{equation}
The Laplace transforms are also moment generating functions, hence
expanding 
for small $s$
\begin{equation}
\hat{\psi}(s) = 1 - s \langle \tau \rangle + s^2 {\langle \tau^2 \rangle \over 2}
+ \cdots
\label{eqM04}
\end{equation}
where $\langle \tau^m \rangle$ are the moments of the times between resets
and similarly
\begin{equation}
\hat{f}_{{\rm ssB}} (s) = 1 - s \langle B \rangle_{{\rm ss}} + s^2 {\langle B^2 \rangle_{{\rm ss}} \over 2} + \cdots .
\label{eqM05}
\end{equation} 
Inserting Eqs. (\ref{eqM04},\ref{eqM05}) in Eq. 
(\ref{eqM03}) we find
 $$ 1 - s \langle  B \rangle_{{\rm ss}} + s^2{ \langle B^2 \rangle_{{\rm ss}}\over 2} + \cdots =  $$
\begin{equation}
{ 1 - \left( 1 - s \langle \tau \rangle + s^2 {\langle \tau^2 \rangle\over 2} - {s^3 \langle \tau^3 \rangle \over 3!} + \cdots \right) \over
s \langle \tau \rangle}.
\label{eqM06}
\end{equation}
Comparing terms of the same order we have $\langle B \rangle_{{\rm ss}}= \langle \tau^2 \rangle / (2 \langle \tau \rangle)$, $\langle B^2 \rangle_{{\rm ss}} = \langle \tau^3 \rangle/ (3 \langle \tau \rangle) $ and in general
$\langle B^m\rangle_{{\rm ss}} = \langle \tau^{m +1} \rangle / [ (m+1) \langle \tau \rangle]$. 
From Eq. (\ref{eqM01}) we find
\begin{equation}
\langle x^2\rangle_{{\rm ss}} = D { \langle \tau^2 \rangle \over \langle \tau \rangle}  
\label{eqM07}
\end{equation}
and more generally
\begin{equation}
\langle x^{2 m} \rangle _{{\rm ss}} = { 4^m  \over (m+1) } { \Gamma\left( m + {1\over 2} \right) \over \sqrt{\pi}} { D^m \langle \tau^{m+1} \rangle \over \langle \tau \rangle}.
\label{eqM08}
\end{equation}

\subsection{Sharp restart is the squeezed state}

 A natural question is to what extent can we squeeze the steady state distribution of $x$? Of course fast resetting will simply put the particle always
on the origin. If we fix $\langle \tau \rangle$ to some non-zero value,
the narrowest steady state § PDF of $x$, will be naively found when $\psi_{{\rm det}}(\tau) = \delta( \tau - \langle \tau \rangle)$. This strategy is called sharp restart and its NESS was
studied in \cite{Eule2016}.
Further, sharp restart is optimal for search \cite{Pal2017,Eliazar2020}
hence studied extensively \cite{Yin}. 
In this case the variance of $x$, is $\langle x^2 \rangle_{{\rm ss}}= D \langle \tau \rangle$ which is smaller that any other $\langle x^2 \rangle_{{\rm ss}}$
found with another  choice of
$\psi(\tau)$, since in general  $\langle \tau^2 \rangle/\langle \tau \rangle \ge \langle \tau \rangle$ 
and hence sharp restart gives the minimum of the dispersion. 

\subsection{No NESS for Sharp restart}

However, sharp restart implies that
the PDF $\psi(\tau)$ being non-smooth is in fact not in the domain of
the theory discussed here, as we mentioned from the start. 
For sharp restart there is
no NESS. To see this note that at any time $t$ slightly larger then 
an integer times $\langle \tau \rangle$, the particle is on $x\simeq 0$.
In contrast just before these times, the PDF of $x$
is a Gaussian with variance, $ 2 D \langle \tau \rangle$. In short,
for stroboscopic resetting, that starts at time $t=0$, we have no
NESS, in the sense that the PDF of $x$ is time dependent, with a periodicity
which is the sharp time between resets. 
 One can claim that if the restart is nearly sharp, i.e.
if we have a narrow but analytical PDF of $\psi(\tau)$ around some   
$\langle \tau \rangle$, namely some small uncertainty in the resetting times,
the process will converge to the NESS. While this is correct,
this convergence will take very long, and the narrower is the
PDF of reset times, around the sharp reset time $\langle \tau\rangle$, 
the longer
will the relaxation towards the NESS will be. 
Another option of obtaining a NESS, for lattice PDFs of resetting time,
is to randomize the initial clock, however this option is not part of this
work.

\subsection{Fat-tailed distributions}

Eq. (\ref{eqM01}) is still valid when $0<\alpha<1$. Noticing that 
$B^m$ is non-integrable with respect to the infinite density
Eq. 
(\ref{eq07}),
since the later decays like $B^{-\alpha}$ for large $B$, we  realize
that the moments are determined by the large $x$ behavior of the propagator
$\rho(x,t)$,
when the scaling presented in Eq. 
(\ref{eq20})
 is diffusive. One may in principle find the moments $\langle x^{2 m}(t) \rangle$  using the properties of the  Kummer function,
however there is no need for that.  Eq. 
(\ref{eqM01}) is still valid and the moments increase with time,
for example 
\begin{equation}
\langle x^2(t) \rangle \sim 2 D \langle B(t)\rangle_{{\rm Dyn}}.
\label{eqM09}
\end{equation}
In turn the moments of the backward recurrence time are obtained using  Dynkin's
limiting law Eq. 
(\ref{eq08},\ref{eq09}). For $m=0,1,..$ we use  
\begin{equation}
\int_0 ^{1} {y^{m-\alpha} \over (1- y)^{1-\alpha}}
{\rm d} y = { \Gamma(\alpha) \Gamma(m-\alpha+1) \over m!},
\label{eqM10}
\end{equation}
and hence
\begin{equation}
\langle B^m (t) \rangle_{{\rm Dyn}} \sim {  \Gamma(m-\alpha+1) \over m! \Gamma(1-\alpha)} t^m.
\label{eqM11}
\end{equation}
Then we find the diffusive type of scaling for the moments
\begin{equation}
\langle x^{2 m} (t) \rangle \sim D^m \langle \xi^{2 m} \rangle \langle B^m(t) \rangle_{{\rm Dyn}}
\label{eqM12}
\end{equation} 
which is made explicit with Eqs. (\ref{eqM02N},\ref{eqM11}).
In the limit $\alpha<<1$ we have $\langle B^m (t) \rangle_{{\rm Dyn}}  \sim t^m$ since then the resetting is extremely  sparse, and as expected
the moments $\langle x^{2 m} (t)\rangle$ are determined by free diffusion.

 We see that the moments of the process are determined by the scaling solution
of $B$ Eq. (\ref{eq08}), 
and hence are not sensitive to the details of the waiting time
PDF $\psi(\tau)$. This is because the moments explore the large $x$ part of the
density $\rho(x,t)$. Another class of observables, considered in the next
section, are integrable with respect to the non-normalized state. And these
do not exhibit diffusive scaling, and their ergodic properties are  of special
interest as they are related to the non-normalized NESS found here. 

\section{Fractional Integral Equation for the density}
\label{SecUni}

The goal of this section is to find a valid approximation for  $\rho(x,t)$ in the limit of long times, this should hold both for large and small $x$.
As we showed  the infinite density approach works well for small $x$, the scaling solution works well for large $x$, so now we want to marry the two solutions, using a uniform approximation. Further, close to the transition $\alpha=1/2$ the convergence to asymptotic results is extremely slow, and this can be overcome with the uniform approximation.

%%%%%%%%%%
\begin{figure}[htbp]
	\includegraphics[width=1\hsize]{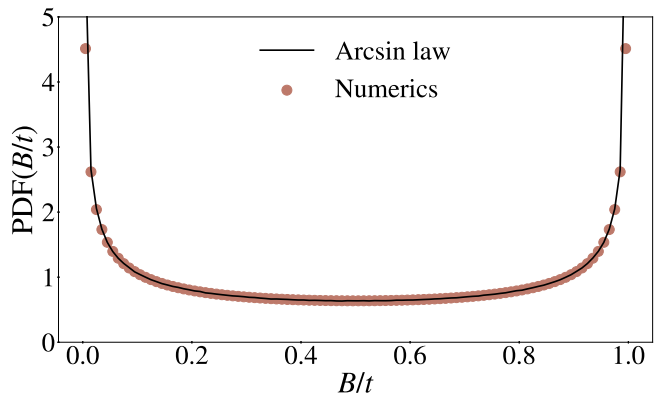}
	\caption{ Numerical simulations for the PDF of the rescaled backward recurrence time $0<B/t<1$ for $\alpha=1/2$ converge to the arcsine law. 
Deviations from this well known  behavior are presented in figure \ref{fig5}}
	\label{fig4}
\end{figure}
%%%%%%%%%%%%%

We focus on fat tailed PDFs of waiting time
Eq. (\ref{eq04}),
$\psi(\tau) \sim (\tau_0)^\alpha \tau^{- 1 -\alpha}$, and $0<\alpha<1$.
The Laplace transform of this function, for small $s$ is provided by
\cite{WeiWang2021,Godreche2001,Klafter}
\begin{equation}
    \hat{\psi}(s) \sim 1 - b_\alpha s^\alpha + \cdots, \; \text{where}\;\; b_{\alpha}=(\tau_0)^{\alpha}\Gamma (1-\alpha)/\alpha
    \label{106A}
\end{equation}

 We start with a recap of a handful of known results 
from the field of renewal processes 
\cite{Godreche2001,Wanli2018,Cox1962}. 
 
%%%%%%%%%%
\begin{figure}[htbp]
	\includegraphics[width=1\hsize]{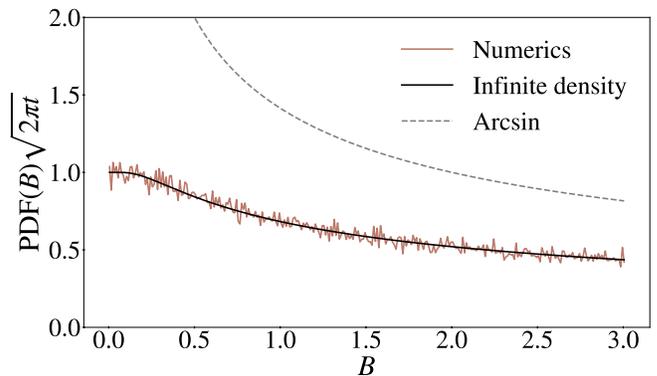}
	\caption{ Numerical simulations for the PDF of the backward recurrence time $B$, multiplied by $\langle \tau^*(t) \rangle$, perfectly match the prediction based on
the infinite invariant  density 
Eq. (\ref{AppendixA05}).
Note that
 $0<B<3$ while $t=10^4$ hence the figure presents the
small $B$ behavior of the density. Here $\alpha=1/2$ hence the arcsine law
Eq.
(\ref{eq09})
 describes the typical fluctuations of $B$, however as shown and as expected it 
fails here (it works fine in the large $B$ regime).
As a stand alone the arcsine law predicts that the density will diverge 
when $B \to 0$, which is clearly not the case for any finite  $t$. 
In simulations  we use $10^7$ realizations of the renewal process. The PDF of waiting times  is given in Eq.
(\ref{eqSmir}) hence $\langle \tau^{*}(t)= \sqrt{ 2 \pi t}$.}
	\label{fig5}
\end{figure}
%%%%%%%%%%%%%

\subsection{Statistics of number of jumps}

Let $P_t(N)$ be the probability for $N$ renewals in the period $(0,t)$.
Using the convolution theorem of Laplace transform
\begin{equation}
\hat{P}_s (N) = { 1 - \hat{\psi}(s) \over s} \left[ \hat{\psi}(s)\right]^N
\label{AppendixAPNS01}
\end{equation}
where $\hat{P}_s(N)$ is the Laplace $t \to s$ transform of $P_t(N)$.
The mean number of jumps is $\langle N(t) \rangle$ and its
 Laplace pair reads
\cite{Shlesinger1974}
\begin{equation}
\langle \hat{N}(s) \rangle = \sum_{N=0} ^\infty N \hat{P}_s(N)=
{\hat{\psi}(s) \over s \left[ 1 - \hat{\psi}(s) \right]}.
\label{AppendixA01}
\end{equation}
Inserting Eq.
(\ref{106A}) and inverting to the time domain 
\begin{equation}
\langle N(t) \rangle \sim { \sin \pi \alpha \over \pi} \left( { t \over \tau_0} \right)^\alpha 
\label{AppendixA02}
\end{equation}
where we used the reflection formula for the $\Gamma$ function.
Eq. (\ref{AppendixA02}) obeys \eqref{eq12aaa}.

\subsection{Backward time statistics}

  A technique 
for finding the distribution of the backward recurrence time is given 
in \cite{Godreche2001}, and it is based on double Laplace transforms.
Let $f_B(t,B)$ be the PDF of $B$ and $\hat{f}_B(s,u)=
\int_0 ^{\infty} {\rm d} t  \int_0 ^{\infty} {\rm d} t\exp( - s t - u B) f_B (t,B)$.
Without any approximation 
\begin{equation}
\hat{f}_B(s,u) = { 1 - \hat{\psi}(s+u) \over s + u} { 1 \over 1 - \hat{\psi}(s) }.
\label{AppendixA03}
\end{equation}

Here $\hat{\psi}(s+u)= \int_0 ^\infty \exp[ - (s+u)\tau] \psi(\tau) {\rm d} \tau$. In principle, if we can invert this formula to the double time domain,  i.e. $t$ and $B$, we can find $\rho(x,t)$ using Eq. (\ref{eq03}).
Using Eq. 
(\ref{106A}), in the limit when $s \to 0$ and $u \to 0$ their ratio remaining finite
\begin{equation}
\hat{f}_B (s,u) \sim s^{-\alpha} (s + u)^{\alpha-1}.
\label{AppendixA04}
\end{equation}
This equation is independent of the fine details of the waiting time PDF besides $\alpha$. The inversion to the time domain is carried out in
\cite{Godreche2001} yielding Dynkin's limit theorem Eqs.
(\ref{eq08},\ref{eq09}). If the mean waiting time is finite,
one uses Eq. (\ref{AppendixA03}) to find Eq. 
(\ref{eq06}). Technically this is done by considering the limit $s \to 0$ while
leaving $u$ fixed, which in turn,  upon Laplace inversion, gives the long time limit of the problem.

 Switching back to $0<\alpha<1$ such that the mean waiting time diverges, we consider Eq. 
(\ref{AppendixA03}) in the limit $s\to 0$ and $u$ finite \cite{Wanli2018}.
Using Eq. (\ref{106A})
\begin{equation}
\hat{f}_B (s,u) \sim { 1 - \hat{\psi}(u) \over u} { 1 \over b_\alpha s^\alpha} .
\label{AppendixA04}
\end{equation}
Inverting to the $(t,B)$ domain one finds Eq. 
(\ref{eq07}).
This describes the statistics of finite $B$ when $t$ is made large.
 
\subsection{Example $\alpha=1/2$}

 To demonstrate this behavior we consider an example. 
Let 
\begin{equation}
\psi(\tau) = { \exp\left( - {1 \over  2 \tau} \right) \over \sqrt{ 2 \pi} \tau^{3/2}}
\label{eqSmir}
\end{equation}
 hence
$\alpha = 1/2$. This PDF is called the one sided L\'evy
 stable distribution with index $1/2$, and is known as a van der Waals profile. 
 In this case $\sqrt{\tau_0} = \sqrt{ 2/\pi}$ and
 $\langle \tau^* (t) \rangle = \sqrt{2 \pi} t^{1/2}$.
 Using 
Eq. (\ref{eq07}) we find
\begin{equation}
\lim_{t \to \infty} \langle \tau^{*} (t) \rangle f(B,t)= \mbox{Erf}\left( {1 \over \sqrt{2 B}} \right),
\label{AppendixA05}
\end{equation}
where we introduced the error function. 
Recall that $\mbox{Erf}(1/\sqrt{ 2 B})  \sim 1$ for $B\to 0$ and
 $\mbox{Erf}(1/\sqrt{ 2 B})  \sim \sqrt{2/\pi} B^{-1/2}$ for large $B$.
Hence the $B$ integration of this infinite invariant density diverges 
$\int_0 ^{\infty} \mbox{Erf} ( 1/ \sqrt{ 2 B} ) {\rm d} B= \infty$, 
due to the large $B$ limit.

 We now consider finite time simulations to demonstrate the result.
Generating the sequence of waiting times
we find the statistics of  $B$ using $10^7$ samples. The random
 waiting times
are given by $\tau=1/G^2$, where $G$ is a Gaussian random variable
with zero mean, whose PDF is  $\exp[ - G^2/2] / \sqrt{ 2 \pi}$ 
\cite{Chambers}.
Generating such a normally distributed random variable with a computer program
is a standard routine.
Hence it is easy to generate the realizations of the
renewal sequence and sample the random variable $B$ on a computer. 

 In Fig. \ref{fig4} we plot the typical fluctuations of $B$ which are  
captured by Dynkin's
limit theorem, in fact since $\alpha=1/2$ we find the arcsine law.
To do so we plot the histogram of the random variable $B/t$ which is clearly
bounded in the unit interval. One sees the well known $U$ shaped histogram, meaning that small $B$ and large $B$ are by far more likely if compared
to the mean which in the long time limit is $\langle B / t \rangle = 1/2$.
Small deviations from the asymptotic theory  are observed on the left,
and those are rare events.

 To study these we focus on $0<B<3\ll t$. Here the infinite density
is a valid approximation while the arcsine law is clearly invalid. 
In Fig. 
(\ref{fig5}) we plot the 
normalized histogram for the backward time, namely the 
sample estimation of the PDF of $B$, multiplied by 
$\langle \tau^{*} (t) \rangle$ versus $B$.
The numerical result matches $\mbox{Erf}(1/\sqrt{ 2 B})$ without any fitting,
indicating that also for finite time simulations the non-normalized result
is a good approximation.

 %%%%%%%%%%
\begin{figure}[htbp]
	\includegraphics[width=1\hsize]{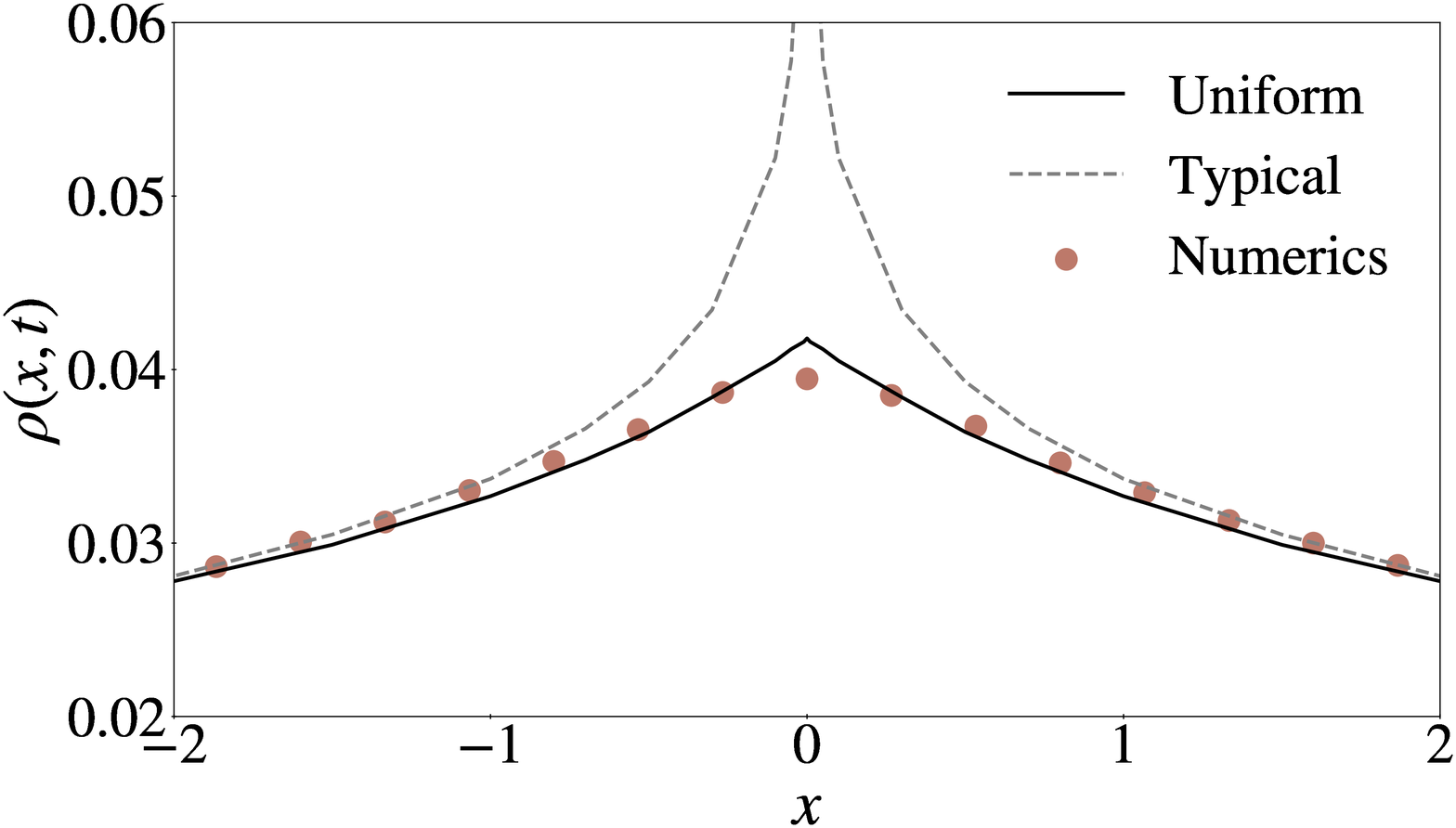}
	\caption{ 
Numerics for the PDF of the position $x$ for an ensemble of
particles undergoing
the  resetting process with $\alpha=1/2$. 
We focus on the small $x$ behavior, namely the vicinity of the resetting point
$x=0$. 
The uniform approximation
Eq. (\ref{eqPXuni})
 perfectly matches the simulations, while the
scaling solution Eq. (\ref{eqPXArcsine}),
 describing the typical fluctuations (large $x$) fails.
Here
$t=10^3$,  
the resetting time PDF is Eq.
(\ref{eqSmir}) and we used
$10^7$ realizations.
}
	\label{fig6}
\end{figure}
%%%%%%%%%%%%%

\subsection{Uniform approximation}

 We have considered already the PDF of $B$, $f_B(B,t)$ in two limits. The typical behavior Eq.
(\ref{eq08}),
  when
$B$ is scaled with time $t$ and the rare events
Eq. 
(\ref{eq07}).
An important  scale of  $\psi(\tau)$, 
 is roughly speaking the time beyond which 
$\psi(\tau) \sim (\tau_0)^\alpha \tau^{-1 - \alpha}$ is a valid approximation. 
For the Pareto distribution
Eq. (\ref{eqppaa})
this time scale is $t_0$ while for the one sided L\'evy stable law 
Eq. (\ref{eqSmir}) it is of order unity. 
For $B$ larger than this time scale the two solutions match, as mentioned
already.

Now we present a simple uniform approximation for the density of
$B$
\begin{equation}
f_{{\rm Uni}} (B,t) = \left\{
\begin{array}{c c}
S(B) \over \langle \tau^{*} (t - B) \rangle & \ 0<B<t \\
0 & \mbox{otherwise}.
\end{array}
\right.
\label{eqUniform}
\end{equation}
 This is obtained by matching the above mentioned two solutions.
Eq. (\ref{eqUniform}) holds for large $t$. 
By construction, for large $B$, we have 
$S(B) \sim (\tau_0/B)^\alpha/\alpha$ and hence this solution matches
Dynkin's limit theorem Eq. (\ref{eq08}), while for small $B$ it yields 
Eq. (\ref{eq13})

 We find the uniform approximation for the density of resetted particles. Using
Eqs. 
(\ref{eq03},
\ref{eqUniform})
\begin{equation} 
\rho_{{\rm Uni}} (x,t) = \int_{0} ^t {\rm d} B { S(B) G(x,B) \over \langle \tau^{*}(t- B)\rangle } .
\label{eqUUNNII}
\end{equation}
This is one of the main results of the paper as it provides both the large $x$ limit of the density $\rho(x, t ) $ (described also by the scaling solution) and the small $x$ limit (given by the infinite density). 
Employing Eq.
(\ref{eq12})
for
$\langle \tau^{*}(t)\rangle$ and the definition of the left sided
fractional Riemann-Liouville integration 
\begin{equation}
_0 D_t ^{-\alpha} g(t) = {1 \over \Gamma(\alpha)} \int_0 ^{t} g(t') (t - t')^{\alpha-1} {\rm d} t' \end{equation}
we find
\begin{equation} 
 (\tau_0)^\alpha |\Gamma(-\alpha)|  \rho_{\rm Uni} (x,t) =
\ _{0} D_t ^{ -\alpha} \left[ S(t) G(x,t) \right].
\label{fe}
\end{equation}
This equation holds far beyond the case of Brownian motion. It connects the survival probability,
the propagator for reset free motion, and the density of the spreading particles. 
It describes both the small $x$ limit which is dominated by small $B$ statistics, as well
as the scaling solution to the problem, discussed previously. 
Note that the inverse operation of the fractional integration is a fractional derivative,
hence we may use Marchaut's formula to find,  at least in principle,
 the product $S(t) G(x,t)$, from  the density
of an ensemble of particles undergoing the resetting process.

%%%%%%%%%%
\begin{figure}[htbp]
	\includegraphics[width=1\hsize]{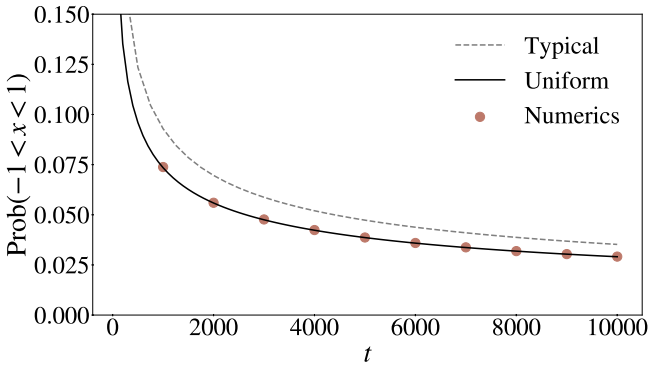}
	\caption{ 
The probability of occupying the interval $-1<x<1$ for a resetting process
to the origin $x=0$ versus time. Here we use the one sided L\'evy distribution Eq. \eqref{eqSmir} 
to model the  time intervals between resetting, so $\alpha=1/2$ which is the transition exponent. Simulations nicely match the uniform approximation
Eq.
(\ref{Eqprob11})
while the theory based on typical fluctuations 
Eq. (\ref{eqpm1x1typ})
does not work so well.
Here $D=1/2$ and for simulations we use $2\cdot 10^7$ trajectories.}
	\label{fig7}
\end{figure}
%%%%%%%%%%%%%

\subsection{The transition case $\alpha=1/2$}

Using $\alpha=1/2$ as an example and for the waiting time PDF Eq. 
(\ref{eqSmir})
we have from Eq. (\ref{eq14})
  the survival probability  $S(B)=\mbox{Erf}(1/ \sqrt{ 2 B})$, 
and $\langle \tau^*(t) \rangle = \sqrt{ 2 \pi t}$, as mentioned.
Therefore the uniform approximation reads
\begin{equation}
\rho_{{\rm Uni}} (x,t) = \int_{0} ^t {\rm d} B { \mbox{Erf}\left({1 \over \sqrt{2 B}}
\right) G(x,B) \over \sqrt{ 2 \pi (t- B) } },
\label{eqPXuni}
\end{equation} 
where we used $D=1/2$. 
This for large $x$ describes well the typical fluctuations given by the arcsine
law for $B$
\begin{equation}
\rho_{{\rm Arcsine}} (x,t) = \int_{0} ^t {\rm d} B { G(x,B) \over \pi  \sqrt{B} \sqrt{t-B} },
\label{eqPXArcsine}
\end{equation}
which is the same as Eq. \eqref{eq17} for $\alpha=1/2$. This approximation diverges on the origin $x=0$ namely in the vicinity
of the resetting point, while in reality and according to Eq.
(\ref{eqPXuni}) such a behavior is not found. 
The two solutions are used in Fig. 
\ref{fig6}
where the  numerics clearly demonstrates that
the uniform approximation is the valid theory. 
The uniform solution and numerics exhibit a cusp in the density close to $x=0$.

Using Eq. (\ref{eqPXuni}) we obtain the probability of finding the particles
in the interval $(-1,1)$. We use 
$\int_{-1} ^{1} \exp[ - x^2/(2 B)] {\rm d x} /\sqrt{ 2 \pi B} = \mbox{Erf}[1/ \sqrt{ 2 B} ]$ to find
\begin{equation}
\mbox{Prob}_{\rm Uni}  (-1<x<1) = { 1 \over \sqrt{ 2 \pi}} \int_{0} ^{t} { \mbox{Erf}^2 \left[ {1 \over \sqrt{2 B} } \right] \over \sqrt{t - B}} {\rm d} B.
\label{Eqprob11}
\end{equation}
This integral is solved numerically and 
compared with Monte-Carlo simulations in Fig.
\ref{fig7} showing the validity of the uniform
approximation. 
We compare this solution to the one obtained using the description
of the typical fluctuations namely using the scaling function
$g_\alpha(\xi)$ with $\alpha=1/2$. Recall that this solution does not depend
explicitly on the waiting time PDF $\psi(\tau)$ 
besides $\alpha$ of course, unlike the uniform approximation.  
Using Eq. 
(\ref{eq21}) and $D=1/2$
$$ \mbox{Prob}_{{\rm typical}}(-1<x<1) \simeq $$
$$
-\int_{-1} ^{1} {\rm d} x 
 { 1 \over \sqrt{ 2 t} \pi^{3/2}} \left[ 2 \ln |x|/\sqrt{t} - \gamma - 3 \ln 2\right] = $$
\begin{equation}
 {\sqrt{2} \over \sqrt{ t \pi^3}} \left( \ln t + 2 +  \gamma + 3 \ln 2\right).
\label{eqpm1x1typ}
\end{equation} 
This solution is plotted in Fig. 
\ref{fig7} where its performance compared with the uniform approximations
are shown to be weak.  
The $\ln(t)$ logarithmic behavior in \eqref{eqpm1x1typ} indicates the particularly slow 
nature of convergence to asymptotic results, strengthening the need for the uniform approximation at this transition case $\alpha=1/2$.
We also integrated  Eq. 
(\ref{eqPXArcsine}) in the interval $-1<x<1$ to estimate
$\mbox{Prob}(-1<x<1)$. This solution works slightly better
than the simple analytical expression Eq.
(\ref{eqpm1x1typ}) yet still not matching the uniform approximation
Eq. 
(\ref{Eqprob11}). It is not plotted to avoid burdening the eye.

\subsection{Uniform Approximation: An Example} 

\begin{figure}[htbp]
	\includegraphics[width=1\hsize]{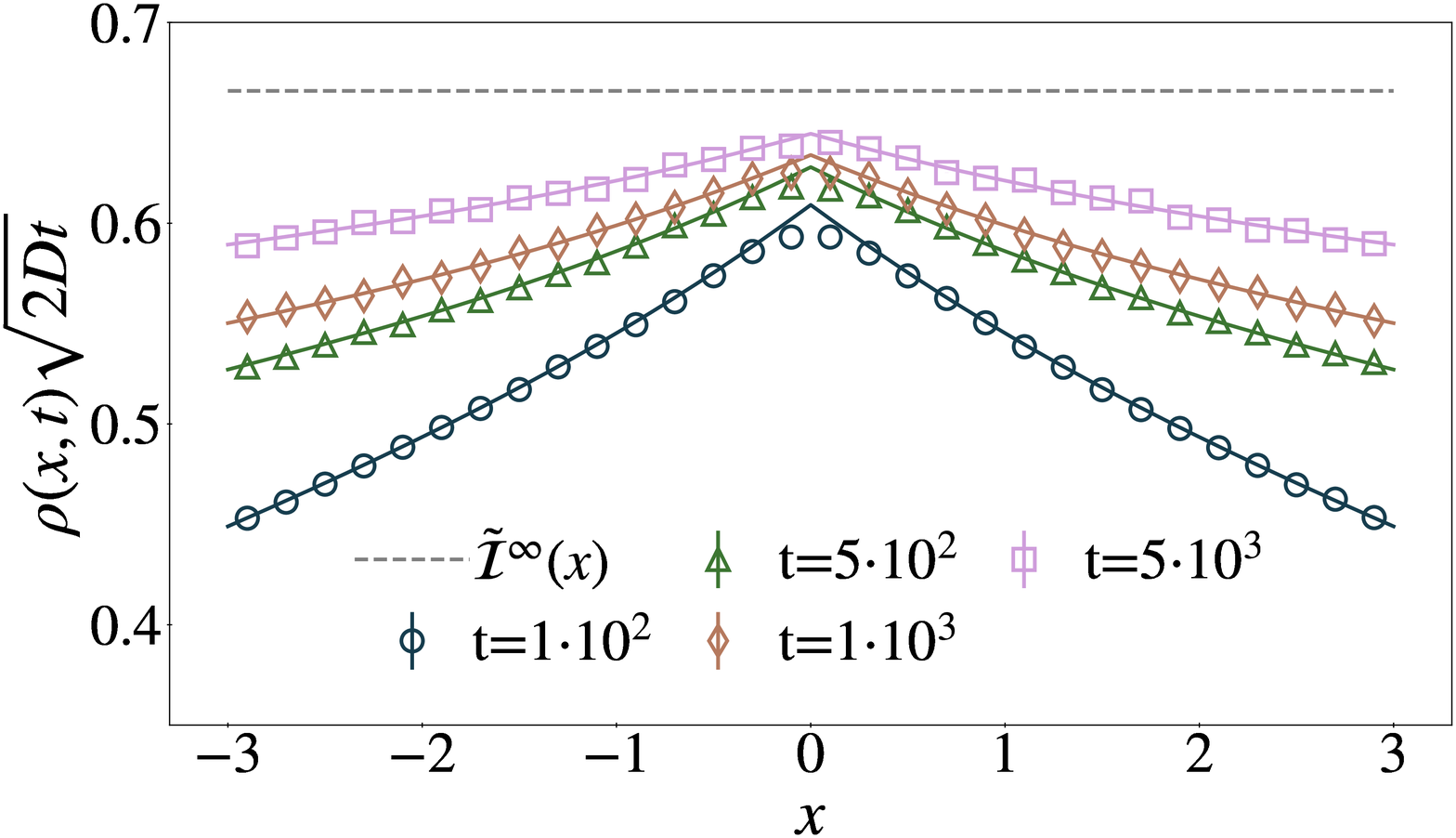}
	\caption{ The infinite density versus $x$ for using the Pareto PDF for waiting times with $\alpha=1/4$ and $t_0 = 1$.
In simulation we plot the histogram for the
probability density $\rho(x,t)$ multiplied by $\sqrt{2 D t}$. In solid lines the uniform approximation Eq. (\ref{equn14}) for intermediate times and in dashed lines the infinite density $\tilde{{\cal I}}^\infty(x)\simeq 0.6659$, Eq. (\ref{eq22}). Here $D=1/2$ and we use $10^8$ trajectories.}
	\label{fig8}
\end{figure}

We now check the predictions of the uniform approximation
using the Pareto distribution for waiting times and  $\alpha=1/4$ 
so
$\psi(\tau) = 0.25 \tau^{-5/4}$ for $\tau>1$, otherwise $\psi(\tau)=0$.
In this case, from Eq. 
(\ref{eq04}),
 $(\tau_0)^{1/4}=1/4$
and using Eq. 
(\ref{eq12}) $\langle \tau^*(t) \rangle = \sqrt{2} \pi t^{3/4}$.
The survival probability is $S(B)= 1$ if $B<1$ and $S(B)=B^{-1/4}$ for $1<B$
hence using 
Eq. (\ref{eqUUNNII}) we have for $t>1$
\begin{equation}
\rho_{{\rm Uni}} (x,t) = \int_{0} ^{1}  {  G(x,B) {\rm d} B \over \sqrt{2} \pi  (t- B)^{3/4} } + 
\int_{1} ^{t}  {  G(x,B) {\rm d} B  \over \sqrt{2} \pi B^{1/4}  (t- B)^{3/4}} .
\label{equn14}
\end{equation}
This solution should be compared with the one obtained using
the scaling solution  Eq. 
(\ref{eq17})
\begin{equation}
\rho(x,t) \sim {1 \over \sqrt{2} \pi}  \int_{0} ^{t} { G(x,B) {\rm d} B \over 
 (t- B)^{3/4} B^{1/4} }. 
\label{eqrrhh}
\end{equation}
which is valid when $x$ and $t$ are large while $\xi=|x|/\sqrt{t}$ is finite and
we set  $D=1/2$. This gives
according to Eq. 
(\ref{eq18})
 $\rho(x,t)\sim g_{1/4} (\xi) /\sqrt{t}$  where the scaling function
 $g_{1/4}(\xi)$ is presented
in Eq.
(\ref{eq20}). Recall that from the infinite density 
Eq. (\ref{eq22})
we have
the approximation, valid for finite $x$ and large $t$
\begin{equation} 
\rho(x,t) 
\sim {g_{1/4} (0)\over \sqrt{2 D t}} 
= 
{ 
\Gamma(1/4) \sqrt{t}  \over \sqrt{2} \Gamma(3/4) \pi}\simeq { 0.66593 \over \sqrt{t}}.
\label{eqin14}
\end{equation}
In Fig. \ref{fig8} we make the comparison between the various approximations. The figure shows that for finite time $t$, the uniform approximation works very well. Of course for the very long time limit the simulations results  will converge to  the theoretical prediction of the infinite density, which is given by the straight line in the Figure.

%%%%%%%%%%
%%%%%%%%%%%%%

We also find the probability that the particle is in the interval
$-1<x<1$ at time $t$.
Using $\int_{-1} ^{1} G(x,B) {\rm d} x = \mbox{Erf}[ 1/\sqrt{ 2 B}]$ we have for the uniform approximation
$$ \mbox{Prob}_{{\rm Uni}} (-1<x<1) = $$ 
\begin{equation}
\int_{0} ^{1} { \mbox{Erf} \left[ { 1 \over \sqrt{ 2 B}} \right] {\rm d} B 
\over \sqrt{2} \pi (t - B)^{3/4}  }
+
\int_{1} ^{t} { \mbox{Erf}\left[ {1 \over \sqrt{ 2 B}} \right] {\rm d} B \over
\sqrt{2} \pi \left( t - B\right)^{3/4} B^{1/4} } . 
\label{eq:probuni}
\end{equation}
These integrals can be numerically computed using programs like Mathematica or Maple.
On the other hand 
from Eq. 
(\ref{eqin14}) we have
$$ \mbox{Prob} (-1<x<1) \simeq 0.66592 |\Delta x|/\sqrt{t}$$
 where $|\Delta x|=2$ is the length of the interval $-1<x<1$
under study.

\begin{figure}[htbp]
	\includegraphics[width=1\hsize]{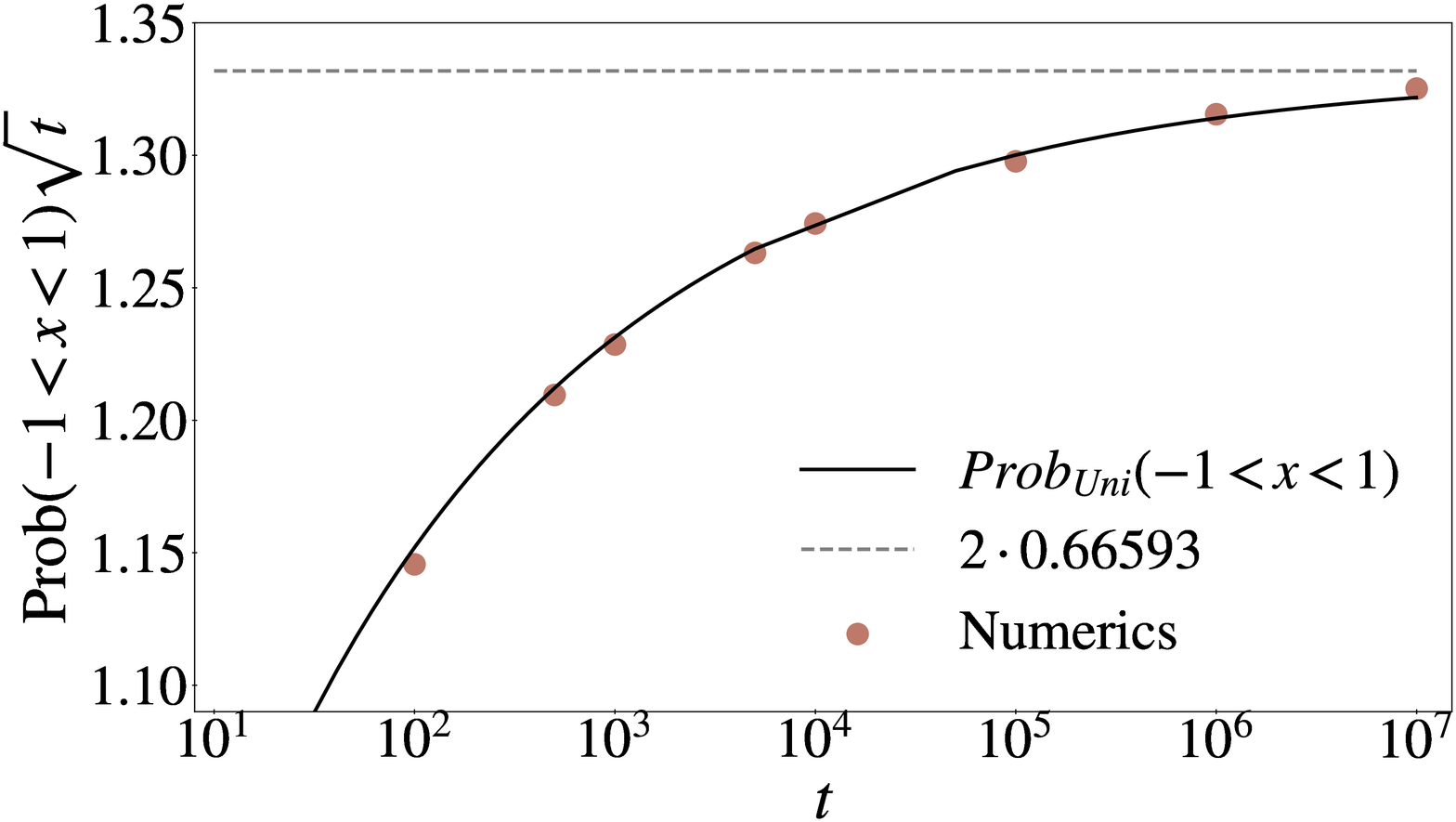}
	\caption{Probability of occupying the interval $-1<x<1$ times $\sqrt{t}$ as a function of time. The points belong to the data from Fig. \ref{fig3}  and Fig. \ref{fig8} while the solid line represents the uniform approximation Eq. (\ref{eq:probuni}). Here the times between resets follow a Pareto PDF with $t_0 = 1$, $\alpha=1/4$. $D$ is $1/2$ and we use $10^8$ realizations.  }
	\label{fig9}
\end{figure}

Fig. \ref{fig9} clearly  demonstrated the useful aspect of the uniform approximation as it captures the approach to the asymptotic limit, i.e. the straight line in the figure.

%%%%%%%%%%
%%%%%%%%%%%%%

\section{Ergodic theory}

 So far we have studied useful approximation for the density of particles. We now study the ergodic properties  of the process. Consider an observable, namely  a functional of the stochastic
 path of the resetting process, ${\cal O}[x(t)]$ 
\cite{Meylahn2015,Hollander2019}.
The time averages are denoted by
\begin{equation}
\overline{{\cal O}(t)} =\frac{1}{t} \int_0 ^{t} {\cal O}[x(t')] {\rm d} t' 
\label{eqEr01}
\end{equation}
while the ensemble average is 
$\langle {\cal O}(t) \rangle = \int_{-\infty} ^\infty {\cal O}(x) \rho(x,t) {\rm d} x$. For thin tailed PDFs of resetting times,
time and ensemble averages
are identical in the long time limit 
\begin{equation}
\lim_{t \to \infty} \overline{{\cal O}(t)}  = \langle {\cal O} \rangle^{{\rm ss}} 
\label{eqEr02}
\end{equation}
and
\begin{equation}
\langle {\cal O}\rangle^{{\rm ss}} = \int_{-\infty} ^\infty {\cal O}(x) \rho^{{\rm ss}}(x){\rm d} x
\label{eqEr03}
\end{equation}
where  
the normalized NESS density $\rho^{{\rm ss}} (x)$ is defined in Eq.
(\ref{eq15}).
We did not prove this expected
result, but some arguments why it is correct are given below. 
Note that in  Eq. (\ref{eqEr03}) we have assumed that the integral 
on the right hand side does not diverge, namely that the observable
${\cal O}(x)$ is integrable with respect to the normalized steady state.
So far in this section  $\langle \tau \rangle$ was finite, 
what is the ergodic theory when this mean time diverges? 

 For fat tailed resetting times, with $0<\alpha<1$
infinite ergodic theory holds. 
This means that the non-normalized steady state will play a special role
in the evaluation of the time averages. First consider the ensemble averages,
in the long time limit we find two types of behaviors. For $1/2<\alpha<1$
using Eq. (\ref{eq16})
\begin{equation} 
\langle {\cal O} (t) \rangle = \int_{-\infty} ^\infty {\cal O}(x) \rho(x,t) {\rm d} x  \sim { \int_{-\infty} ^\infty {\cal O}(x) \tilde{{\cal I}}^\infty (x) {\rm d} x \over \langle \tau^{*}(t) \rangle}.
\label{eqEr04}
\end{equation}
where we use Eq. (\ref{eq16}), so  $\rho(x,t) \sim \tilde{{\cal I}}^\infty(x)/\langle \tau^{*}(t) \rangle$.
Similarly using Eq. (\ref{eq22}) for $0<\alpha<1/2$ 
\begin{equation}
\langle {\cal O} (t) \rangle \sim {\int_{-\infty} ^\infty {\cal O} (x) \tilde{{\cal I}}^\infty (x) {\rm d} x \over \sqrt{ 2 D t} }.
\label{eqEr05}
\end{equation}
We assumed that the integrals do not diverge, namely that the observable is integrable with respect to the infinite invariant density
 $\tilde{{\cal I}}^\infty(x)$. 
Eqs. (\ref{eqEr04},\ref{eqEr05}) show that while $\tilde{{\cal I}}^{\infty}(x)$ is not
normalized, it is used to obtain ensemble averages.  
More precisely
\begin{equation}
\begin{array}{l l}
\lim_{t \to \infty} \langle \tau^* (t) \rangle \langle {\cal O}(t) \rangle =
\int_{-\infty} ^\infty {\cal O}(x) \tilde{{\cal I}}^\infty (x) {\rm d} x \  & \mbox{if} \  1/2 < \alpha<1 \\
\ & \ \\
\lim_{t \to \infty} \sqrt{2 D t}  \langle {\cal O}(t) \rangle =
\int_{-\infty} ^\infty {\cal O}(x) \tilde{{\cal I}}^\infty (x) {\rm d} x \ & \mbox{if} \  0 < \alpha<1/2,
\end{array}
\end{equation}
thus $\langle \tau^* (t) \rangle$ and $\sqrt{ 2 D t}$ replace normalizing factors.

 An example for an integrable observable consider
\begin{equation} 
{\cal O}[x(t)] = \theta(a<x(t)<b)
\label{eqEr06}
\end{equation}
where $\theta(a<x(t)<b)$ is the pulse function, namely it is  equal unity if the condition in the parentheses holds, otherwise it is zero. 
Since the integral $\int_{a} ^b\tilde{ {\cal I} }^\infty (x){\rm d} x$
 is finite, for finite $a$ and $b$, the observable is called integrable, we will now use
this observable to discuss time averages.  

\subsection{Example}

\begin{figure}[htbp]
	\includegraphics[width=1\hsize]{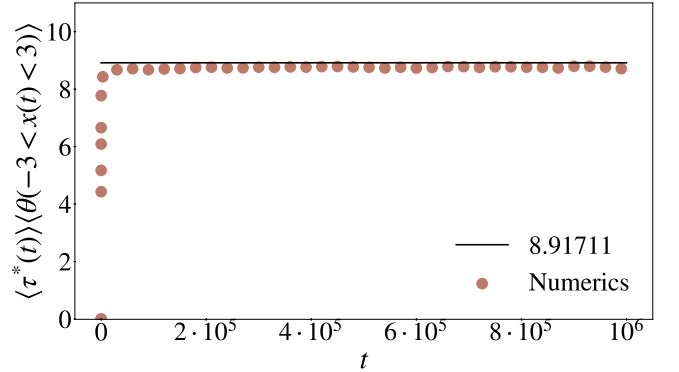}
	\caption{Time evolution of  $\langle \theta(-3<x(t)<3)\rangle$ times $\langle \tau^{*} (t) \rangle$ Eq. 
(\ref{eqttaauu}) for a Pareto PDF with $\alpha=3/4$ and $t_0=1$. The solid line represents the long time limit Eq. (\ref{eq:thetalim}). Here $D=1/2$ and we have used $10^6$ trajectories. }
	\label{fig10}
\end{figure}

As an example consider the case $\alpha=3/4$ with the Pareto PDF discussed
in subsection 
\ref{SecEx} with $D=1/2$. The observable of interest is the pulse function
$\theta(-3<x(t)<3)$. Using the infinite density  Eq. 
(\ref{Infa304})
\begin{equation}
\langle \tau^{*} (t) \rangle \langle \theta(-3<x(t)<3)\rangle\sim  \int_{-3} ^{3} \tilde{{\cal I}} ^{\infty} _{\alpha=3/4} (x) {\rm d} x.
\end{equation}
The integral is solved numerically and we find
\begin{equation}
\lim_{t \to \infty} \langle \tau^{*}(t) \rangle \langle \theta(-3<x(t)<3) \rangle = 8.91711 \cdots
\label{eq:thetalim}
\end{equation}
and $\langle \tau^{*} (t) \rangle$ is given in Eq. 
(\ref{eqttaauu}) with $\alpha=3/4$. 
As mentioned $\langle \theta(-3<x(t)<3)\rangle$ is the probability that a member of an ensemble of particles  occupies the domain $[-3,3]$ at time $t$.
This prediction is tested in Fig. \ref{fig10} showing that the non-normalized invariant density is the tool of choice to compute ensemble averages of integrable observables.

%%%%%%%%%%
%%%%%%%%%%%%%

\subsection{Time averages}

The time integration over the pulse function observable
Eq. (\ref{eqEr06}) is the total
time a trajectory $x(t')$ spends in the domain $[a,b]$ during the measurement
time interval $(0,t)$, it will be denoted $\tilde{T}$.
 For example $[a,b]$ can be a domain in space including the resetting point, or not. The total time
the particle spends 
in $[a,b]$ is called the occupation time or the residence time 
\cite{Pal2019,Bressloff2020,Singh2022}. For thin
tailed distributions of resetting times, and using the ergodic hypothesis
\begin{equation}
\lim_{t \to \infty} { \int_0 ^t \theta(a<x(t') <b) {\rm d} t' \over t} =
 \int_{a} ^b \rho^{{\rm ss}}(x) {\rm d} x
\label{eqEr06aa}
\end{equation}
which is the probability a member of an ensemble of particles
in NESS occupies the domain. 
We will treat this observable for the case $\alpha<1$ below.

 We now treat an integrable observable not restricting our selves to an example.
Then the ensemble averaged time average 
\begin{equation}
\langle \overline{{\cal O }}(t) \rangle 
\sim {1 \over t} \left\langle \int_0 ^t {\cal O}(t') {\rm d} t' \right\rangle
\end{equation}
 which is found by averaging many of the underlying processes, over time 
and over independent trajectories.  Since the ensemble average in any experiment is simply a sum
over a finite sample, we may
 replace the order of time and ensemble
average and then
\begin{equation}
\langle \overline{{\cal O}}(t) \rangle = {\int_{0} ^{t}\left[ \int_{-\infty} ^\infty \rho(x,t') {\cal O}(x) {\rm d} x \right] {\rm d} t' \over t}.
\label{eqEr07}
\end{equation}
As before, in the long time limit we replace the density $\rho(x,t)$
with the infinite density using Eq.
(\ref{eq16})
, e.g. for $1/2<\alpha<1$ 
\begin{equation}
\langle \overline{{\cal O}}(t) \rangle \sim  { \int_{0} ^{t} \left[ \int_{-\infty} ^\infty    { \tilde{{\cal I}}^{\infty}(x) \over \langle \tau^{*}(t') \rangle } {\cal O}(x)  {\rm d} x \right]  {\rm d} t' \over t}.
\label{eqEr07}
\end{equation}
In the numerator we can identify the ensemble average obtained with integration
over the infinite density Eq. 
(\ref{eqEr04}). Further the time integration is straight forward, using 
Eqs.  
(\ref{eqEr04},\ref{eqEr05})
 we find 
\begin{equation}
\lim_{t \to \infty} {\langle \overline{{\cal O}}(t) \rangle \over \langle {\cal O}(t) \rangle} = 
\left\{
\begin{array}{c c}
1   & \ \mbox{for} \   \psi(\tau) \ \mbox{thin tailed} \\
\ & \\
{1  \over \alpha} & \ {1 \over 2}< \alpha<1  \\ 
\ & \\
2   &  \ 0<\alpha< {1 \over 2}. 
\end{array}
\right.
\label{eqEr08}
\end{equation}
We see that the time and ensemble  averages are related one to another.
For $\alpha<1$ they are calculated using the infinite
density other wise by the normalized invariant density of the NESS.  
The prefactors found for $\alpha<1$ stem from simple time integration.
We see that when $\alpha=1$ and $\alpha=1/2$ there is an ergodic transition
in the system which is in principle easy to detect when $\alpha$ 
is tuned  (the result in the first line, holds for any distribution,
of resetting times, even a distribution that decays like a power law $\alpha>1$,
in such a way that the mean resetting time is finite). The predictions in Eq. \eqref{eqEr08} are tested with finite time simulations, showing excellent agreement without fitting.

%%%%%%%%%%
\begin{figure}[htbp]
	\includegraphics[width=1\hsize]{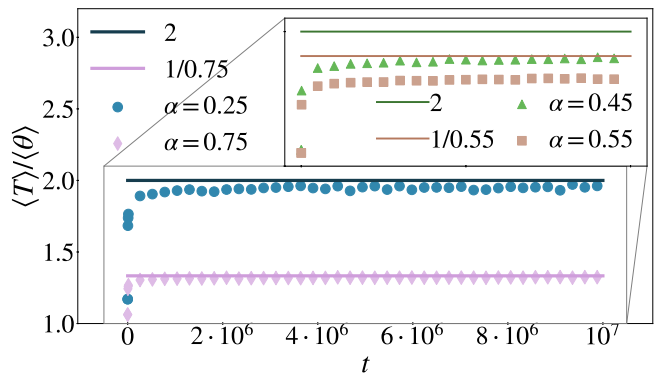}
	\caption{ Ratio between the ensemble-averaged
time average and the ensemble average versus time. The observable is the pulse function Eq. (\ref{eqEr06}) in the interval $-10 < x < 10$ . We present numerical results for the Pareto reset time PDF with $t_0=1$ with $\alpha = 1/4$ and $\alpha = 3/4$. In solid lines the long time limit of $\langle T\rangle /\langle \theta\rangle $, Eq. (\ref{eqEr08}). In the inset figure, we present the same results but for $\alpha$ near the transition point below, $\alpha=0.45$ and above $\alpha=0.55$. We can see that the agreement is not so good due to the slowing down of the convergence near the transition in agreement with what is observed also in Fig. \ref{fig12}. Here $D=1/2$ and we have used $10^7$ trajectories.}
	\label{fig11}
\end{figure}
%%%%%%%%%%%%%

 Infinite ergodic theory deals also 
with the limiting laws of the  distribution of time averages.
 We define a dimensionless variable
\begin{equation}
\eta = {\overline{{\cal O}} \over\langle \overline{{\cal O}} \rangle}
\label{eqEr09}
\end{equation}
and clearly the mean of $\eta$ is unity. 
In what follows assume that the observable is integrable, namely the non-zero denominator is 
obtained in theory from the invariant density
using Eqs. 
(\ref{eqEr04},\ref{eqEr05},\ref{eqEr08})
though a direct measurement 
in say an experiment or simulation
is also a good  possibility. 
 For thin tailed waiting time PDFs, $\eta$ in the long time limit is not fluctuating. In other words its distribution is a delta function centred on unity.

\subsection{Fluctuations of time averages}

Consider the integral over the pulse function
Eq. (\ref{eqEr06})
\begin{equation}
\eta =  {\int_0 ^{t} \theta(-a<x(t)< a) {\rm d} t \over
 \langle \int_0 ^{t} \theta(-a<x(t)< a) {\rm d} t \rangle}= {\tilde{T} \over \langle \tilde{T} \rangle}. 
\label{eqEr12A}
\end{equation} 
 Namely we are interested in the statistics
of the occupation time $\tilde{T}$ in $[-a,a]$ when the observation is in 
$(0,t)$.
Of course the interval  $[-a,a]$
contains the resetting point $x=0$. 
In Eq. (\ref{eqEr12A}) $\eta$ is normalized in the sense that its mean is unity.

The function $\theta(-a<x(t)<a)$ attains the value $1$ when $x(t)$ is
in the domain $[-a,a]$, $0$ other wise.  The observable is switching
at random times between values $1$ and $0$,  namely it is performing
a dichotomous two state process. What is the physical mechanism of the return into the domain $[-a,a]$? One option is that the resetting returns the particle to the domain, namely just before a resetting event the particle is say on
$x(t)> a$ and it is injected back to $-a<x=0<a$. Yet another option
is that the particle, via the process of diffusion alone, returns back into
the domain. We have here a competition between these two mechanisms
of return.  Recall that 
 the PDF of resetting times is given by a fat tailed
law Eq. 
(\ref{eq04}) while the PDF of first passage time of BM in an infinite domain
in dimension one decays like $(\mbox{time})^{-3/2}$
\cite{Redner}, in the absence of
resetting.
Hence we might  expect a transition in the ergodic properties of the system when $\alpha=1/2$, which is also noticed in the behavior of the infinite invariant densities discussed above. 

%%%%%%%%%%
\begin{figure}[htbp]
	\includegraphics[width=1\hsize]{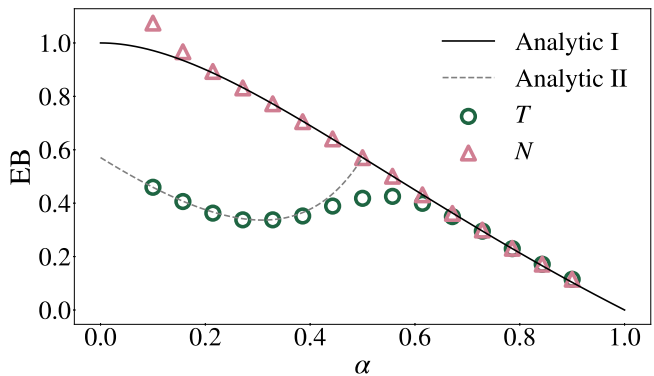}
	\caption{The EB parameter, describing the fluctuations of the time averages, versus $\alpha$. Numerical results for the occupation time $T$, for the reset process are presented, where we used $D= 1/2, a=0.1$ the Pareto reset time PDF,
with $t_0= 1$. For $1/2<\alpha<1$, the EB parameter is the  same as
that for $N$ (triangles), the number of resets up to time $t$. Close to the transition point $\alpha=1/2$ we see deviations from theory, due to finite sampling
and slowing down of the convergence. 
When $0<\alpha<1/2$ the fluctuations are non-trivial and we observe a minimum
for the EB parameter. Here we have used $10^5$ trajectories evolved since $t=10^{10}$. We used Eq. \eqref{eqtttt1} (Analytic II) and Eq. \eqref{eqEEBB} (Analytic I) to present the theory.}
	\label{fig12}
\end{figure}
%%%%%%%%%%%%%

 To quantify this behavior we use the EB parameter \cite{He2008}, defined as
\begin{equation}
\mbox{EB} ={ \langle \eta^2 \rangle - \langle \eta\rangle^2 \over \langle \eta \rangle^2} = {\langle \tilde{T}^2 \rangle -\langle \tilde{T} \rangle^2 \over \langle \tilde{T} \rangle^2 }.
\label{eqtttt}
\end{equation}
An analysis of the EB parameter  starts in the next section and more details are
found in Appendix B and C. 
In the long time limit 
\begin{equation}
\textrm{EB}=\left\{ \begin{array}{cc}
\alpha\pi-1+\frac{1}{2}\left[\frac{\pi\Gamma\left(1-\alpha\right)}{\Gamma\left(\frac{1}{2}-\alpha\right)}\right]^{2}, & 0<\alpha<1/2\\
\ & \ \\
\frac{2\Gamma^{2}(1+\alpha)}{\Gamma(1+2\alpha)}-1, & 1/2<\alpha<1 \\
\ & \ \\
0 , & \mbox{thin tailed PDFs} 
\end{array}\right.
\label{eqtttt1}
\end{equation}
When the mean of the resetting time PDF is finite there is no ergodicity breaking as the PDF of $\eta$ converges to a delta function and $\rm{EB}=0$. In contrast if $0<\alpha<1$ we have two types of behaviors. Note that
when $\alpha\to 0$, we have $\mbox{EB} = (\pi/2) -1\simeq 0.57$.
As shown below, just after Eq. 
(\ref{eqBMO4a})
this is the value of the EB parameter for free BM (see also Appendix B).  

 When $1/2<\alpha<1$ the fluctuations we find are related to the fluctuations
of the number of resets in the time interval $(0,t)$.
More specifically recall that
$N$ is the random number of resets in the time interval
$(0,t)$. The EB parameter for this observable is
well known \cite{He2008}
\begin{equation}
{\rm EB}= { \langle N^2 \rangle - \langle N \rangle^2 \over \langle N \rangle^2}=
\frac{2\Gamma^{2}(1+\alpha)}{\Gamma(1+2\alpha)}-1,
\label{eqEEBB}
\end{equation}
which is valid for $0<\alpha<1$ and in the long time limit. 
We see that the fluctuations of the time averages are related to the fluctuations of the number of resettings, however this is true only when $1/2<\alpha<1$.
The intuitive explanations is that when $1/2 <\alpha$  the return to the domain
$[-a,a]$ is dominated by resetting and not by diffusion. Further the time
spent in side the domain $[-a,a]$ is statistically short compared
to the time outside the domain, and hence it is the fluctuations of the latter 
that dominate the statistics of the time averages. The same is not 
true for thin tailed PDFs where the mean times in and out side of 
the domain are both finite. Hence in the latter case we find ordinary ergodicity.

To summarize, one can say that for $0<\alpha<1/2$ the fluctuations are
less trivial if compared to other cases. 
As mentioned more technical  details are provided below.
In Fig. \ref{fig12} we present numerical results for the EB parameter.
We observe a minimum for the EB parameter found also in
the context of a study of laser cooling \cite{Barkai2021}.
The convergence of finite time simulations is poor close
to the critical value of $\alpha=1/2$, an effect that could be studied further.

\begin{widetext}
\section{Occupation time statistics}

As mentioned,  we call the time spent by the resetting process $x(t')$
  in the spatial domain
$[-a,a]$, within the time window $(0,t)$, $\tilde{T}$, where the resetting is
to the origin $x=0$. 
The PDF of this
random variable will be denoted $P(\tilde{T},t)$. 
For standard ergodic processes, and in the limit of long times, we expect
that this PDF becomes a narrow distribution, centred around the mean,
as mentioned already.
However, when $\alpha<1$ this is not true any more. To analyze this
issue, we use a tool developed by Montroll, Weiss and others
 in the context of continuous time random walks \cite{Klafter}. 
The same tool was used to study ergodic properties of sub-recoil laser cooled
gases \cite{Barkai2021}. 
Here the first  goal is to relate between statistics of occupation times of Brownian
motion and those of the occupation times of the reseted process. 
Secondly, we derive the basic formulas of infinite ergodic theory from
a well known approach, and further provide simple intuitive formulas for
averages. We start with a recap of occupation time statistics for Brownian motion.

\subsection{Occupation time for Brownian motion}

 Consider reset free Brownian motion $x_{{\rm BM}}(t)$ starting on
the origin $x=0$.  The occupation time of the process is
\begin{equation}
T_{{\rm BM}}(t) = \int_{0} ^{t} \theta[x_{{\rm BM}}(t')] {\rm d} t'
\label{eqBMO1}
\end{equation}
Here and in what follows  we use $\theta[x_{{\rm BM}}  (t)] = \theta( -a <x_{{\rm BM}}(t)<a)$ as a short hand notation. 
The mean of the pulse function is
obtained from the Gaussian packet
\begin{equation}
\langle \theta[x_{{\rm BM}}(t)] \rangle =
\int_{-\infty} ^\infty  G(x,t) \theta[x] {\rm d} x. 
\label{eqBMO2}
\end{equation}
Hence the mean occupation time is  $\langle T(t)\rangle_{{\rm BM}}= \int_{0} ^t \langle \theta[x_{{\rm BM}}(t') ] \rangle {\rm d} t'$ and using Eq. (\ref{eq01})
 it is easy to show that
\begin{equation}
\langle  T(t) \rangle_{{\rm BM}} =
t + { a \sqrt{t} \over \sqrt{ \pi D}} \exp\left( - {a^2 \over  4 D t} \right)
- \left(t + {a^2 \over  2 D} \right) \mbox{Erfc}\left( { a \over \sqrt{ 4 D t}} \right).
\label{eqBMO3}
\end{equation}
In the long time limit
\begin{equation}
\langle T (t) \rangle_{{\rm BM}} \sim { 2 a \over \sqrt{ \pi D}} t^{1/2}.
\label{eqBMO4}
\end{equation}
The distribution of the occupation time of Brownian motion is discussed
further in Appendix B. Using the Feynmann-Kac formalism \cite{Kac,Functionals},
 the PDF of the
occupation time in the long time limit is half a Gaussian
\begin{equation}
\mbox{PDF}_{{\rm BM}}(T|t) \sim \left( { D \over \pi  a^2 t} \right)^{1/2} 
 \exp\left( - { D T^2 \over 4 a^2 t} \right).
\label{eqBMO4a}
\end{equation} 
The Laplace transform of the 
finite time solution is presented in Appendix B.
The long time limit of the  EB parameter is $\mbox{EB} = ( \langle T^2 \rangle_{{\rm BM}} - \langle T\rangle_{{\rm BM}} ^2 )/ 
\langle T\rangle_{{\rm BM}} ^2  = (\pi/2)-1$, namely the same as
the reseted process Eq.
(\ref{eqtttt1})
 when $\alpha \to 0$.

\subsection{Occupation time for the resetting problem}

The occupation time of the reseted process 
$\tilde{T}(t)= \int_{0} ^{t} \theta[x(t')] {\rm d} t'$ is now considered.
With the notation for time averages  Eq.
(\ref{eqEr01})
and  for the pulse function observable  $\overline{{\cal O}} = \tilde{T}/t$.
 Let $Q_N ( \tilde{T}, t) {\rm d} t {\rm d} \tilde{T}$ be the probability that the $N$th resetting event takes place in the time interval $(t, t+ {\rm d} t)$ when the value of the occupation time is within $(\tilde{T}, \tilde{T} + {\rm d} \tilde{T})$.  This function is given by the iteration rule
\begin{equation}
Q_{N+1} (\tilde{T},t) = \int_{0} ^{\tilde{T}} {\rm d} T' \int_{0} ^{t} {\rm d} \tau Q_N \left( \tilde{T} - T' , t - \tau\right) \phi(T',\tau).
\label{eqOT01}
\end{equation}
Here $\phi(T',\tau)$ is the joint PDF 
of the resetting interval $\tau$  i.e. the time between consecutive resets, 
and the occupation time in the same interval $T'$.
According to Eq. (\ref{eqBack}) the resetting process is defined with a sequence of time
intervals between resettings
$$ \left\{ \tau_1 , \tau_2, ... \tau_N, B\right\}.$$
Within each resetting interval $\tau_i$ we have an occupation time
$T_i$ in the spatial  domain $(-a,a)$. Given the reset time interval $\tau_i$,
statistical properties of $T_i$ are determined by the laws of Brownian motion.
 Clearly the occupation
time for the reseted process is
\begin{equation}
\tilde{T} = \sum_{i=1} ^{N} T_i + T_B.
\label{eqOT02}
\end{equation}
$T_B$ is the occupation time in the interval $(-a,a)$ gained
in the backward time $B$. The sets $\{ T_i \}$ and $\{ \tau_i \}$ are separately 
composed from IID random variables, however $T_i$ and $\tau_i$ are mutually
dependent. The longer is $\tau_i$ the longer $T_i$ is, in statistical sense.
The joint PDF of the pair is 
\begin{equation}
\phi(T, \tau) = \psi(\tau) \mbox{PDF}_{{\rm BM}}(T | \tau).
\label{eqOT03}
\end{equation}
Where here we use the PDF of occupations times for Brownian motion without
restarts. 

Eq. (\ref{eqOT01}) describes the basic property of the process.
To arrive in $\tilde{T}$ at time $t$ when the previous resetting
took place at $t-\tau$, the previous value of $\tilde{T}$, at the moment
of the previous reseting, was $\tilde{T} - T$. Notice that in 
Eq. (\ref{eqOT01}) time $t$ denotes a dot on the time axis, on which a
resetting took place (see details below).
 Solving Eq. 
 (\ref{eqOT01}) is possible with the help of the convolution
theorem of Laplace transform.
Let 
\begin{equation}
\hat{Q}_N (p,s) = \int_{0} ^{\infty} {\rm d} \tilde{T} \int_{0} ^\infty {\rm d} t \exp\left( - p \tilde{T} - s t \right) Q_N (\tilde{T}, t )
\label{eqOT04}
\end{equation}
be the double Laplace transform of $Q_N (\tilde{T}, t)$ where 
$p \leftrightarrow \tilde{T}$ and $s \leftrightarrow t$ are Laplace 
pairs.
The convolution theorem and the iteration rule give
$\hat{Q}_{N+1} (p,s) = \hat{Q}_{N} (p,s) \hat{\phi}(p,s)$ where
$\hat{\phi}(p,s)$ is the double Laplace transform of $\phi(T,\tau)$.
Using the seed, $Q_0(\tilde{T},t) = \delta(\tilde{T})\delta(t)$,
reflecting the initial condition, namely that the resetting process starts
at time $t=0$, we have
\begin{equation}
\hat{Q}_N (p,s) =  \left[ \hat{\phi}(p,s) \right]^N.
\label{eqOT05}
\end{equation}
The PDF $P(\tilde{T},t)$ is in turn given by
\begin{equation}
P(\tilde{T}, t) = \sum_{N=0} ^\infty
\int_{0} ^{\tilde{T} } {\rm d} T^{'} _B \int_{0} ^t {\rm d} B 
Q_N\left( \tilde{T} - T^{'} _B, t - B \right) \Phi(T_B , B). 
\label{eqOT06}
\end{equation}
Here we summed over the number of restarts $N$ and took into consideration
the fact that the observation time $t$, is found at  a time
$B$  after the last resetting
event in the sequence. We further integrate over the backward recurrence time.
Finally, the statistical weight function is 
\begin{equation}
\Phi(T_B, B) =  S(B) \mbox{PDF}_{{\rm BM}} (T_B | B) 
\label{eqOT07}
\end{equation}
where as before $S(B)= \int_{B} ^\infty \psi(\tau) {\rm d} \tau$ is the survival
probability, i.e. the probability of not resetting. We will soon use the Laplace
$B \to s$ transform of this function $\hat{S}(s)=  [ 1 - \hat{\psi}(s)]/s$ 
where $\hat{\psi}(s)$ is the Laplace transform of $\psi(\tau)$.  

 Now we again use the convolution theorem. Let $\hat{P}(p,s)$ be the double
Laplace transform of $P(\tilde{T},t)$. Using Eq. 
(\ref{eqOT06}) 
\begin{equation}
\hat{P}(p,s) = \sum_{N=0} ^\infty \hat{Q}_N (p,s) \hat{\Phi}(p,s)
\label{eqOT08}
\end{equation}
where
\begin{equation}
\hat{\Phi}(p,s) = \int_{0} ^\infty \int_{0} ^\infty \exp[ - T_B p - B s] 
S(B) \mbox{PDF}_{{\rm BM}} (T_B | B ) {\rm d} T_B {\rm d} B. 
\label{eqOT09}
\end{equation}
Inserting Eq. (\ref{eqOT05}) 
Eq. (\ref{eqOT08})
and summing the geometric series
\begin{equation}
\hat{P}(p,s) = { \hat{\Phi}(p,s) \over 1 -\hat{\phi}(p,s)}.
\label{eqOT10}
\end{equation}
In the context of continuous time random walks such an equation is used
to analyse the positional PDF of the packet of particles, though then
usually one invokes a Fourier-Laplace transform \cite{Klafter,Fifty,Sergey,Aghion}. The inversion of
the formal solution Eq. (\ref{eqOT10}) to the $(\tilde{T},t)$ domain is
a significant problem, which can be tackled analytically in the long time limit. In particular from the definition of the Laplace transform
$$
\hat{P}(p,s) = \int_{0} ^\infty {\rm d} \tilde{T} \int_{0} ^\infty {\rm d} t \exp[ - p \tilde{T} - s t] P\left( \tilde{T}, t\right) = $$
$$ \int_{0} ^{\infty} ( 1 - p \tilde{T} + \cdots) {\rm d} \tilde{T} \int_{0} ^\infty
{\rm d} t e^{- s t}  P(\tilde{T}, t)$$
$$  = {1 \over s} - p \langle \tilde{T}(s) \rangle \cdots$$
where we used the normalization condition for $P(\tilde{T},t)$ and
$\langle \hat{\tilde{T}}(s) \rangle$ is the Laplace $t \to s$ transform of
the mean occupation time $\langle \tilde{T}(t) \rangle$. Another way
to write this is
\begin{equation}
\langle \hat{\tilde{T}}(s) \rangle = - { \partial \hat{P}(p,s) \over \partial p}|_{p=0}.
\label{eqOT11}
\end{equation}
Using the Montroll-Weiss like equation (\ref{eqOT10}) we find
\begin{equation}
\langle \hat{\tilde{T}}(s) \rangle =
- { \partial_p \hat{\Phi}(p,s)|_{p=0}
\over 1 - \hat{\phi}(p=0,s) }  -
{\partial_p \phi(p,s)|_{p=0} \hat{\Phi}(p=0,s) \over
[ 1 - \phi(p=0,s)]^2}. 
\label{eqOT12}
\end{equation}
Similar approach can be used to find the variance of the occupation time, however
here will will study the mean only. 

 We use $\hat{\phi}(p=0,s) = \hat{\psi}(s)$, $\hat{\Phi}(p=0,s)= [1 -\hat{\psi}(s)]/s$, and 
\begin{equation}
\partial_p \hat{\phi}(p,s)|_{p=0} = - \int_{0} ^\infty \exp(-s \tau) \langle T(\tau) \rangle_{{\rm BM}}  \psi(\tau) {\rm d} \tau
\label{eqOT12}
\end{equation}
where 
\begin{equation}
\langle T(\tau) \rangle_{{\rm BM}} = \int_{0} ^\infty T \cdot  \mbox{PDF}_{{\rm BM}} (T|\tau)  {\rm d} T
\label{eqOT13}
\end{equation}
is the mean occupation time of a Brownian motion in $(-a,a)$, in the time
interval $(0,\tau)$. Similarly 
\begin{equation}
\partial_p \hat{\Phi}(p,s)|_{p=0} = - \int_{0} ^\infty \exp(-s B) \langle T(B) \rangle_{{\rm BM}}  S(B) {\rm d} B. 
\label{eqOT14}
\end{equation}
Using Eq. (\ref{eqOT12}) we find
\begin{equation}
 \langle \hat{\tilde{T}}(s) \rangle= 
\underbrace{ { \int_{0} ^\infty \langle T(\tau) \rangle_{{\rm BM}} \psi(\tau) e^{- s \tau} {\rm d} \tau \over s \left[ 1 - \hat{\psi}(s) \right]}}_{ {\cal T}_1(s) } 
+ 
\underbrace{
{ \int_{0} ^{\infty} \langle T(B) \rangle_{{\rm BM}} S(B) e^{ - s B} {\rm d} B  \over
1 - \hat{\psi}(s) }
}_{{\cal T}_2(s)}.
\label{eqOT14}
\end{equation}
This formula relates between the mean occupation time
of the resetting process, with the mean occupation time
of the restart free Brownian motion, and with the waiting
time $\psi(\tau)$. 
We note that Eq.  (\ref{eqOT14}) can be generalized to other observables
beyond the occupation time.
The two contributions defined in this equation, namely 
${\cal T}_1(s)$ and
${\cal T}_2(s)$, describe contributions to the occupation time before
and after the last reset event in the sequence.

\subsubsection{ Mean occupation time  $ 1/2<\alpha<1$}

 To analyze the long time behavior of the mean occupation time we
consider the small $s$ limit, following a standard approach by considering Eq. \eqref{106A} which holds for $0<\alpha<1$. 
We now need to distinguish between three cases.
A short calculation, valid when  $1/2<\alpha<1$ will convince the reader
that  the leading term, when $s\to 0$ in Eq. 
(\ref{eqOT14}) reads
\begin{equation}
\langle \hat{\tilde{T}}(s) \rangle  \sim { \int_{0} ^{\infty} \langle T(\tau) \rangle_{{\rm BM}} \psi(\tau) {\rm d} \tau \over b_\alpha s^{1 + \alpha} },
\label{eqOT15}
\end{equation}
and  only ${\cal T}_1(s)$ is contributing to this limit. 
Inverting to the time domain we find in the long time limit
\begin{equation}
\langle \tilde{T}(t) \rangle \sim { t^\alpha
   \over \Gamma(1 + \alpha) b_\alpha}
 \int_{0} ^{\infty} \langle T(\tau) \rangle_{{\rm BM}} \psi(\tau) {\rm d} \tau.
\label{eqOT15}
\end{equation}
Noticing that the average number of restarts is $\langle N(t) \rangle_{{\rm Res}} \sim
t^\alpha /b_\alpha \Gamma(1 + \alpha)$ we have
\begin{equation}
\langle \tilde{T}(t) \rangle \sim \langle N(t) \rangle_{{\rm Res}} \langle \langle T \rangle_{{\rm BM}} \rangle_{{\rm Res}}
\label{eqOT16}
\end{equation}
where the mean of the occupation time, within a resetting period, averaged over
the resetting time is
\begin{equation}
\langle \langle T \rangle_{{\rm BM}} \rangle_{{\rm Res}} =
\int_{0} ^\infty \int_{0} ^\infty T \cdot \mbox{PDF}_{{\rm BM}} ( T| \tau ) {\rm d} T \psi(\tau) {\rm d} \tau.
\label{eqOT17}
\end{equation}
In Eqs. (\ref{eqOT16},\ref{eqOT17}) 
we distinguish between averages over the reseting time, and the averages over the Brownian motion within each interval. Eq. (\ref{eqOT16})
is expected, 
a main point to notice is that it is not valid when $0<\alpha<1/2$. 
Since  $\langle T(\tau) \rangle_{{\rm BM}} \sim \tau^{1/2}$, when averaging
over $\psi(\tau) \propto \tau^{-1 - \alpha}$, the integral in Eq.  
(\ref{eqOT17}) diverges when $\alpha<1/2$, a case soon to be treated.

Eq. (\ref{eqOT16})  remains valid for
the case where the mean of the waiting time between resets is finite,
for example when $\psi(\tau)$ is an exponential function. The difference
is that $\langle N(t) \rangle_{{\rm Res}}  \sim t /\langle \tau \rangle$,
where $\langle \tau \rangle$ is the mean time between restarts.

How
is  Eq. (\ref{eqOT15}) related to  the non-normalised
 invariant density when $1/2<\alpha<1$? 
Using the pulse function
Eq. (\ref{eqEr06})
 of a Brownian path without resetting
\begin{equation}
\langle T(\tau) \rangle_{{\rm BM}} =\left \langle \int_{0} ^\tau \theta[x(t')] {\rm d} t' \right\rangle_{{\rm BM}}= 
\int_{0} ^\tau \langle \theta[x(t')] \rangle {\rm d} t'=
\int_{0} ^\tau \int_{-\infty} ^\infty \theta[x] G(x,t') {\rm d} x {\rm d} t'
\label{eqOT18}
\end{equation}
as mentioned already.
Using Eq. 
(\ref{eqOT16}) 
 the occupation time  for the resetting process
is
\begin{equation}
\langle \tilde{T}(t) \rangle \sim \langle N(t) \rangle_{{\rm Res}} 
\int_{0} ^\infty {\rm d} \tau  \int_{-\infty} ^{\infty} {\rm d} x  G(x,\tau) \theta(x) [- \partial_\tau S((\tau)]
\label{eqOT19}
\end{equation}
where we apply $\psi(\tau) = - \partial_\tau S(\tau)$. 
Integrating by parts, and employing Eq.
(\ref{eq16})
\begin{equation}
\langle \tilde{T}(t) \rangle \sim \langle N(t) \rangle_{{\rm Res}} 
\int_{-\infty} ^\infty \theta(x) {\cal I}^\infty (x) {\rm d} x.
\label{eqOT20}
\end{equation}
which for the observable of interest, namely the pulse function
reads
$\langle \tilde{T}(t) \rangle \sim \langle N(t) \rangle_{{\rm Res}} 
\int_{-a} ^a {\cal I}^\infty (x) {\rm d} x$.
By definition $\langle \tilde{T} (t) \rangle /t = \langle \overline{\theta}[x(t)]\rangle$ and hence the CTRW approach and Eq. (\ref{eqOT20}) yield the same result as in Eq. 
(\ref{eqEr08}) utilizing Eqs.
(\ref{eq12aaa},
\ref{eqEr04}) and the long time identity
$\partial_t \langle N(t)\rangle_{{\rm Res}} /\alpha= \langle N(t) \rangle_{{\rm Res}}/t$.

\subsubsection{ Mean occupation time  $ 0<\alpha<1/2$}

  We now analyse the case $0<\alpha<1/2$. Here contributions to the mean occupation
time stem from both terms in Eq. 
(\ref{eqOT14}), namely now the backward recurrence time is large in statistical sense,
in a way that it contributes to the averaged observable also in the long time
limit.
In the small $s$ limit, we use the asymptotic formula Eq.  (\ref{eqBMO4}) and find
employing Eq. (\ref{eqOT14}) and $1-\hat{\psi}(s) \sim b_\alpha s^\alpha$
\begin{equation}
{\cal T}_1 (s) \sim { 2 a \over \sqrt{\pi D} }
{  \int_{0} ^\infty \tau^{1/2} (\tau_0)^\alpha  \tau^{-1 - \alpha} \exp( -s \tau) {\rm d}\tau \over
b_\alpha s^{1+ \alpha} },
\label{eqOT21}
\end{equation}
where we used Eq.
(\ref{eq04}). Inserting the definition of $b_\alpha$
given in  
Eq. (\ref{106A})
and integrating we find for small $s$
\begin{equation} 
{\cal T}_1 (s) \sim { 2 \alpha  a \over \Gamma(1 -\alpha)  \sqrt{ \pi D}} \Gamma \left(1/2-\alpha\right) s^{-3/2}.
\label{eqOT22}
\end{equation}
Inverting to the time domain  we find
\begin{equation}
{\cal T}_1 (t) \sim { 4 a \over \pi \sqrt{ D}} { \Gamma( 1/2 - \alpha) \over  |\Gamma(- \alpha)|}
t^{1/2}
\label{eqOT23}
\end{equation}
The second contribution is analysed similarly, 
in particular employing  Eq.  (\ref{eqBMO4})
\begin{equation}
{\cal T}_2 (s) \sim { 2 a \over \sqrt{ \pi D} } { \int_{0} ^{\infty} \sqrt{B} S(B) \exp( - s B) {\rm d} B \over b_\alpha s^\alpha}. 
\label{eqOT24}
\end{equation}
Using $S(B) \sim (\tau_0)^\alpha \tau^{-\alpha} /\alpha$ obtained from
Eq. 
(\ref{eq04}),
 integrating and then inverting to the time 
domain we find
\begin{equation}
{\cal T}_2(t) \sim { 4 a \over \pi \sqrt{ D} } { \Gamma(3/2- \alpha) \over \Gamma(1-\alpha)} t^{1/2}
\label{eqOT25}
\end{equation}
Summing Eqs. 
(\ref{eqOT23},
\ref{eqOT25}) we get the mean occupation time
\begin{equation}
\langle \tilde{T} (t) \rangle \sim  { 2 a \over \pi \sqrt{D} } { \Gamma({1/2 -\alpha} ) \over \Gamma(1-\alpha) } t^{1/2}.
\label{eqOT26}
\end{equation}
When $\alpha \to 0$ we obtain the same result as found for Brownian motion,
Eq. 
(\ref{eqBMO4}), while in the limit $\alpha\to 1/2$ this expression diverges signalling 
the transition.

The same result can be obtained from the infinite density approach.
Since the occupation time is the time integral of the pulse  function
\begin{equation} 
{ \langle  \tilde{T}(t) \rangle \over t} \sim 2 \langle \theta(x) \rangle
\label{eqOT27}
\end{equation}
where we used Eq. 
(\ref{eqEr08}) which gives the prefactor $2$. 
The average $\langle \theta(x) \rangle$  is with respect to the infinite invariant
density as in Eq. 
(\ref{eqEr05})
\begin{equation}
\langle \theta(x)\rangle \sim { \int_{-\infty} ^\infty \theta(x) {\cal I}^{\infty} (x) {\rm d} x \over \sqrt{ 2 D t} } 
\label{eqOT28}
\end{equation}
As mentioned in Eq.
(\ref{eq22}) the infinite density is a constant in this case. It is then easy to show
that the results obtained with CTRW approach are the same as those found
with the infinite density method. Of course this is what is expected, though
here once we have the infinite density, the calculation is straight forward,
as is the case of ergodic processes, where time integration is replaced with a phase space integration. 
In Appendix B we continue with this line of study and calculate the fluctuations of the time averages, these are needed to obtain the $\rm{EB}$ parameter.

\end{widetext}

\section{Discussion}

 Relating the NESS to the limiting laws of
the backward recurrence times \cite{Godreche2001,Dynkin1955,Wanli2018} 
was our starting point. 
This is a valuable tool for many restart models,  where
the reset erases the memory of the process, and is not limited to BM. 
As studied in \cite{Bordova2019A} 
the  erasure of memory, 
which is
clearly valid for a Markovian BM,
 is not the general rule. 
Using statistics of the backward recurrence time we simplified main
expressions for NESS (that  previously relied on Laplace transforms) and
  obtained results which
were found previously with other methods \cite{Pal2016,Gupta2016,Eule2016}.
We also added new  ingredients to the resetting literature.

 The tools of infinite ergodic theory, and the non-normalised NESS are employed to obtain general ergodic aspects
 of the restart process.  The invariant densities can be normalised or non-normalised,
still their functional  dependence
 on the survival probability  appears similar. Thus controlling the distribution
of resetting time, we can explore either the standard ergodic phase,
or the theory of infinite ergodic theory. The time
scale $\langle \tau^*(t) \rangle$ and the length scale $\sqrt{ 2 D t}$ are
used to relate the infinite density  $\tilde{{\cal I}}^\infty(x)$ with the probability density $\rho(x,t)$, for $1/2<\alpha<1$ and $0<\alpha<1/2$ respectively.
See Eqs. 
(\ref{eq16},\ref{eq22}).

 The behaviors of both time and ensemble averages were addressed.
When dealing with thin tailed  distributions the standard
ergodic picture emerges. The exception is  sharp restart, which has
no NESS. 
The case of sporadic resetting with fat tailed distributed resetting times,
was the main focus of this study. 
A statistical theory of the time averages works as follows. When $\alpha<1$
we first check that the observable is integrable with respect to the non-normalized state. 
In this case we find the ensemble average using the infinite invariant density. 
Once this is known, we use 
Eq. (\ref{eqEr08})
to obtain the ensemble average of the time average  $\langle \overline{{\cal O}}\rangle$.
We then studied the fluctuations of the time averages, focusing on an integrable observable, namely the pulse function. The time integral of this observable is
the total time spent in an interval, called the occupation time.
 The fluctuations exhibited non-trivial effects, and a transition in the EB parameter was found for $\alpha=1/2$.
Additionaly $\alpha=1/2$ marks a transition in the structure
of the infinite invariant density itself.
We further pointed out that when $1/2<\alpha<1$ the EB parameter of the
time average, is the same as the one computed for the fluctuations of the number of renewals. This implies that the fluctuations in this phase are universal,
and independent of the observable, as long as it is integrable,
however we did not prove this statement. In contrast, when $0<\alpha<1/2$ the fluctuations of time averages, and the EB parameter, depend on the observable
and hence non-universal.

We speculate that this type of  transition is generic, and can be found similarly in other processes. 
As we showed, in our case, the transition is found
when  $\alpha$ matches the exponent describing  the PDF
 of first passage times of a Brownian motion on a line in the absence of resetting. The latter well known PDF,
 with absorption at say $x=0$, decays like $(\mbox{first passage time})^{-(1 + \beta)}$ and $\beta=1/2$ in dimension one. 
In many other processes $\beta \neq  1/2$, for example for diffusion
in a potential that grows like the log of the distance, 
or for sub-diffusive CTRW, or random walks on some fractals
 or comb structures etc.  
 We believe that
when $\alpha=\beta<1$ the resetting process might exhibit a transition similar
to what we found here, but the details and the generality of this statement must be worked out. Finally, we have studied the uniform approximation, both for $B$ and for the coordinate $x$ of the reseted particle. This approach gives the  probability  density $\rho(x,t) $ for small and large $x$, and was shown to yield statistical quantities also for intermediate time scales. It tackles the problem of slowing down when $\alpha=1/2$. From this excellent approximation we find the fractional equation \eqref{fe}, which is a simple tool for the calculation of $\rho(x,t)$.

\section{Conclusions}
 We showed how the analysis of the statistics of the backward recurrence time solves the NESS of the restart process. Two types of invariant densities
are present in this problem. These are the normalized and non-normalized  invariant densities,  for thin tailed or fat tailed resetting time PDFs respectively.
We uncovered two ergodic transitions. The first takes place  when the mean waiting time diverges.  The second is found when $\alpha=1/2$. At this critical value of $\alpha$ the infinite density  changes its structure. Further the EB parameters exhibits a non-analytical behavior. Thus both time and ensemble averages have vastly different behaviors when $\alpha<1/2$ compared to $1/2<\alpha<1$.
Physically this ergodic transition is found 
due to the competition between return mechanisms to the origin. We also found
slow convergence to asymptotic limits. 
To tackle this issue we used the uniform approximation. We use a simple fractional integral equation to this end, connecting  fractional calculus to the calculation of the density of particles.

$$ $$
{\bf Acknowledgement}
The support of the Israel Science Foundation (EB) and the Spanish government (RF, VM) under
grants $1614/21$ and PID2021-122893NB-C22, respectively are acknowledged.
\newpage

\appendix
\section{}
\numberwithin{equation}{section}
  If $\langle \tau \rangle$ is finite, our results for the NESS
reduce to those found previously
 by Pal, Kundu, and Evans (PKE)  \cite{Pal2016}.  
PKE consider BM under a time modulated resetting protocol. The rate of resetting
is $r(t)$, and it is a function of time $t$ since the last reset event. 
We fix the resetting position at $x=0$. To see that this model is the same
as the one considered here, 
 we identify $S(\tau)= \exp[ - R(\tau)]$ where $R(\tau)= \int_0 ^{\tau} r(t') {\rm d} t'$, hence the PDF of times between resetting is 
\begin{equation}
\psi(\tau) = -{ {\rm d} \over {\rm d} \tau} S(\tau)= r(\tau)\exp[ - R(\tau)].
\label{eqA01}
\end{equation}
PKE find the NESS using Laplace transforms
\begin{equation}
\rho^{{\rm ss}}(x)= \lim_{s \to 0} {\hat{Q} (x,s) \over \hat{{\cal H}_r} (s) }.
\label{eqA02}
\end{equation}
In the numerator $\hat{Q}(x,s)= \int_0 ^\infty {\rm d} t \exp[ - s t - R(t)] G(x,t)$ and $G(x,t)$ is the Gaussian Green function of the BM Eq.
(\ref{eq01}).
 Since  as mentioned $\exp[- R(t)]$ is the survival probability, $S(t)$ in
our notation, and taking the $s\to 0$ limit, we get 
\begin{equation}
\lim_{s \to 0} \hat{Q}(x,s) = \int_0 ^\infty S(\tau) G(x,\tau) {\rm d} \tau.
\label{eqA03}
\end{equation}
This is the same as the function on the right hand side of  Eq. 
(\ref{eq15}).
Further using PKE's results  
$$\lim_{s \to 0} \hat{{\cal H}_r}(s)=  \int_0 ^{\infty} {\rm d} \tau e^{ - R(\tau)}=  $$
\begin{equation}
\int_0 ^{\infty}{\rm d} \tau S(\tau) = -\int_0 ^{\infty} {\rm d} \tau \tau {{\rm d} S(\tau) \over {\rm d} \tau}= \langle \tau \rangle.
\label{eqA04}
\end{equation}
Hence we see that PKEs result Eq. 
(\ref{eqA02}) is the same as Eq. (\ref{eq15}).

\section{}

 Consider one dimensional Brownian motion starting 
at $x_0$. The PDF of the occupation time, in the spatial domain $(-a,a)$,
is denoted  $\mbox{PDF}_{{\rm BM}} (T|t)$.
 Clearly this PDF is a function of $x_0$ though in main text we study $x_0=0$.
 Let $g_{x_0} (p;t)$ be the Laplace
transform  
\begin{equation}
g_{x_0} ( p ; t) = \int_{0} ^\infty \exp( - p T) \mbox{PDF}_{{\rm BM}}(T|t) {\rm d} T.
\label{eqBMO5}
\end{equation}
The backward  Feynmann-Kac equation 
reads \cite{Functionals,Carmi}
\begin{equation}
\partial_t 
g_{x_0} ( p ; t) = D {\partial^{2} 
g_{x_0} ( p; t)
\over \partial (x_0)^2} 
 - p \theta[x_0] g_{x_0} (p; t).
\label{eqBMO6}
\end{equation}
As well known this is the Schrodinger equation for imaginary time. In this analogy $p \theta[x]$ acts like a potential of force, a square barrier in our case. 
Initially $g_{x_0} (T,t) = \delta(T)$ since the occupation time is zero
at the initial time and hence employing the Laplace transform
we get $g_{x_0} (p;t)|_{t=0} = 1$. 
We now consider a second Laplace transform 
\begin{equation}
g_{x_0} (p;s) = \int_{0} ^\infty \exp( - s t) g_{x_0} (p; t) {\rm d} t.
\label{eqBMO7}
\end{equation}
The variables in the parenthesis of the function define the space we are working in. 
Using Eq. 
(\ref{eqBMO6}) and the initial condition
\begin{equation}
s g_{x_0} (p;s) -1 = D 
 {\partial^{2} 
g_{x_0} (p;s) 
\over \partial (x_0)^2} 
- p \theta[x_0] g_{x_0} (p; s).
\label{eqBMO8}
\end{equation}
Using the pulse function, namely $\theta(x_0)=1$ in $-a<x_0<a$ otherwise zero, 
 we have three regions 
\begin{widetext}
\begin{equation}
g_{x_0}(p;s) = \left\{
\begin{array}{c c}
c_0 \exp \left( { x_0 \sqrt{s} \over \sqrt{D}} \right) + {1 \over s} & \ \ x_0<-a \\
c_1 \exp \left( { x_0 \sqrt{s+p} \over \sqrt{D}} \right)+
c_2 \exp \left( - { x_0 \sqrt{s+ p} \over \sqrt{D}} \right) +  {1 \over s+p } & \ \    -a <x_0<a \\
c_3 \exp\left( - { x_0 \sqrt{s} \over \sqrt{D} } \right) + { 1 \over s} &
\ \ x_0 > a
\end{array}
\right.
\label{eqBMO9}
\end{equation}
Here $c_0,c_1,c_2,c_3$ are constants independent of $x_0$. 
Since $g_{x_0}(p;s) = g_{-x_0}(p;s)$ from
symmetry, $c_1=c_2$ and $c_0=c_3$. We use the boundary at $x_0=-a$ and from the continuity condition
\begin{equation}
c_0 \exp \left(  -{a  \sqrt{s} \over \sqrt{D}} \right) + {1 \over s} =
c_1\left[  \exp \left( -{ a \sqrt{s+p} \over \sqrt{D}} \right)+
 \exp \left( { a \sqrt{s+ p} \over \sqrt{D}} \right) \right] +  {1 \over s+p }.
\label{eqBM10}
\end{equation}
Further from the continuity of the fluxes at the boundary
namely $\partial_{x_0} g_{x_0}(p;s)|_{x_0=-a-\epsilon}=
\partial_{x_0} g_{x_0}(p;s)|_{x_0=-a+\epsilon}$ when $\epsilon\to 0$,
we get
\begin{equation}
c_0 \sqrt{s} \exp\left( - {a \sqrt{s} \over \sqrt{D} }\right)
= -2 c_1 \sqrt{s + p}  \sinh\left( { a \sqrt{ s+p} \over \sqrt{D} } \right).
\label{eqBM11}
\end{equation}
Solving and setting $x_0=0$, which greatly simplifies the solution,  we find
\begin{equation}
g_0 (p;s) = { p \over s (s + p) \left[
\cosh\left( {a \sqrt{ s + p} \over \sqrt{D}} \right) + \sqrt{ 1 + { p \over s} }
\sinh \left( { a \sqrt{ s+ p } \over \sqrt{D} } \right) \right] }
+ { 1 \over s + p} .
\label{eqBM11}
\end{equation}
Setting $p=0$ we have $g_0(0,s)=1/s$ which is the normalization condition.
As usual we have the expansion
\begin{equation}
g_0(p;s) \sim { 1 \over s} - p\langle \hat{T}(s) \rangle_{{\rm BM}} + {p^2 \langle \hat{T}^2(s) \rangle_{{\rm BM}} \over 2} \cdots.
\label{eqBM12}
\end{equation}
Hence the small $p$ expansion of Eq. (\ref{eqBM11}) yields the moments
of the occupation time for $x_0=0$. For example
\begin{equation}
\langle T(s) \rangle_{{\rm BM}} = { 1 - \exp\left( - a \sqrt{s} \over \sqrt{D} \right)  \over s^2}.
\label{eqBM13}
\end{equation}
The inverse Laplace transform gives Eq. 
(\ref{eqBMO3}) in the main text.
The second moment $\langle T^2(s) \rangle_{{\rm BM}}$ is similarly found using a program like Mathematica, though the expression is already cumbersome.
Focusing on the long time limit, we consider the small $s$ expansion
and find $\langle \hat{T}^{2} (s) \rangle_{{\rm BM}} \sim
 2 a^2/(D s^2)$ inverting  yields $\langle T^2(t) \rangle_{{\rm BM}} \sim 2 a^2 t/D$ which in turn gives $\mbox{EB} = (\pi/2) -1$ as mentioned in the text. 
Finally, to find the long time limit, we expand the solution for small $s$ and
$p$ finding
\begin{equation}
g_0 (p;s) \sim { \sqrt{  D /a^2} s^{-1/2} \over p + \sqrt{D/a^2} s^{1/2} }
\label{eqBM14}
\end{equation}
Inverting from $p \to T$ and then $s \to t$ we find the half Gaussian
PDF of the occupation time Eq. 
(\ref{eqBMO4a}) in the main text.

% THE NEXT APPENDIX IS TAKEN FROM VICENC WRITE UP
%
%% LyX 2.3.6.2 created this file.  For more info, see http://www.lyx.org/.
%% Do not edit unless you really know what you are doing.
%\documentclass[english]{article}
%\usepackage[T1]{fontenc}
%\usepackage[latin9]{inputenc}
%\usepackage{amstext}
%\usepackage{esint}
%\usepackage{babel}
%\begin{document}
%
% 

\section{}

\subsection{Mean square occupation time}

From the PDF of the occupation time in the Laplace space, the mean
square occupation timemay be found from
\begin{equation}
\left\langle \tilde{T}^{2}(s)\right\rangle =\left[\frac{\partial^{2}\hat{P}(p,s)}{\partial p^{2}}\right]_{p=0}.\label{eq:t2}
\end{equation}
From Eq. (80) we find
\begin{equation}
\frac{\partial^{2}\hat{P}(p,s)}{\partial p^{2}}=\frac{\partial_{pp}\hat{\Phi}}{1-\hat{\phi}}+2\frac{\partial_{p}\hat{\Phi}\partial_{p}\hat{\phi}}{\left(1-\hat{\phi}\right)^{2}}+\frac{\hat{\Phi}\partial_{pp}\hat{\phi}}{\left(1-\hat{\phi}\right)^{2}}+2\frac{\hat{\Phi}\left(\partial_{p}\hat{\phi}\right)^{2}}{\left(1-\hat{\phi}\right)^{3}}\label{eq:d2p}
\end{equation}
and from Eq. (79) the double Laplace transform of $\phi(T,\tau)$
we also get
\begin{equation}
\partial_{pp}\hat{\phi}|_{p=0}=\int_{0}^{\infty}e^{-s\tau}\psi(\tau)\left\langle T^{2}(\tau)\right\rangle _{\textrm{BM}}\equiv I_{2}(s)\label{eq:I2d}
\end{equation}
and
\begin{equation}
\partial_{pp}\hat{\Phi}|_{p=0}=\int_{0}^{\infty}e^{-sB}S(B)\left\langle T^{2}(B)\right\rangle _{\textrm{BM}}\equiv I_{2}^{*}(s).\label{eq:I2ad}
\end{equation}
Defining $\partial_{p}\hat{\phi}|_{p=0}=-I_{1}(s)$ and $\partial_{p}\hat{\Phi}|_{p=0}=-I_{1}^{*}(s)$
and from (\ref{eq:t2}) and (\ref{eq:d2p}) we have
\begin{equation}
\left\langle \tilde{T}^{2}(s)\right\rangle =\frac{I_{2}^{*}(s)}{1-\hat{\psi}(s)}+\frac{2I_{1}(s)I_{1}^{*}(s)}{\left(1-\hat{\psi}(s)\right)^{2}}+\frac{I_{2}(s)}{s\left(1-\hat{\psi}(s)\right)}+\frac{2I_{1}(s)^{2}}{s\left(1-\hat{\psi}(s)\right)^{2}}\label{eq:t22}
\end{equation}
and the mean occupation time given in Eq. (86) can be rewritten as
\begin{equation}
\left\langle \tilde{T}(s)\right\rangle =\frac{I_{1}^{*}(s)}{1-\hat{\psi}(s)}+\frac{I_{1}(s)}{s\left(1-\hat{\psi}(s)\right)}\label{eq:t1}
\end{equation}

\subsection{Exponential resetting}

Let us obtain first the mean occupation time and the mean square occupation
time for exponential resetting. Considering $\psi(\tau)=re^{-rt}$
we find $I_{1}(s)=r\left\langle \hat{T}(s+r)\right\rangle _{\textrm{BM}}$,
$I_{1}^{*}(s)=\left\langle \hat{T}(s+r)\right\rangle _{\textrm{BM}}$,
$I_{2}(s)=r\left\langle \hat{T}^{2}(s+r)\right\rangle _{\textrm{BM}}$,
$I_{2}^{*}(s)=\left\langle \hat{T}^{2}(s+r)\right\rangle _{\textrm{BM}}$.
In consequence, (\ref{eq:t1}) has the form
\[
\left\langle \tilde{T}(s)\right\rangle =\left(\frac{r+s}{s}\right)^{2}\left\langle \hat{T}(s+r)\right\rangle _{\textrm{BM}}.
\]
In the long time limit $s\rightarrow0$
\[
\left\langle \tilde{T}(s)\right\rangle \simeq\frac{r^{2}}{s^{2}}\left\langle \hat{T}(r)\right\rangle _{\textrm{BM}}
\]
so that in the real time
\[
\left\langle T(t)\right\rangle \simeq\left\langle \hat{T}(r)\right\rangle _{\textrm{BM}}r^{2}t=t\left(1-e^{-a\sqrt{r/D}}\right).
\]
Analogously,
\[
\left\langle \tilde{T^{2}}(s)\right\rangle =\left(1+\frac{r}{s}\right)^{2}\left[\left\langle \hat{T^{2}}(s+r)\right\rangle _{\textrm{BM}}+2r\left(1+\frac{r}{s}\right)\left(\left\langle \hat{T}(s+r)\right\rangle _{\textrm{BM}}\right)^{2}\right]
\]
which in the limit $s\rightarrow0$ 
\[
\left\langle \tilde{T^{2}}(s)\right\rangle \simeq2\frac{r^{4}}{s^{3}}\left(\left\langle \hat{T}(r)\right\rangle _{\textrm{BM}}\right)^{2}
\]
so that in the real time
\[
\left\langle T^{2}(t)\right\rangle \simeq t^{2}\left(1-e^{-a\sqrt{r/D}}\right)^{2}.
\]
We see that $\left\langle T^{2}(t)\right\rangle =\left\langle T(t)\right\rangle ^{2}$
so that $\textrm{EB}=0.$

\subsection{Long-tailed resetting}

\subsubsection{Mean occupation time}

Now we consider the long tailed resetting PDFs using
\begin{equation}
\psi(\tau)=\left\{ \begin{array}{cc}
0, & \tau<t_{0}\\
(\tau_{0})^{\alpha}\tau^{-1-\alpha}, & \tau>t_{0}
\end{array}\right.\label{eq:psi}
\end{equation}
and

\begin{equation}
S(\tau)=\left\{ \begin{array}{cc}
1, & \tau<t_{0}\\
(\tau_{0}/\tau)^{\alpha}\alpha^{-1}, & \tau>t_{0}
\end{array}\right.\label{eq:S}
\end{equation}
where $t_{0}=\alpha^{-1/\alpha}\tau_{0}.$ We compute the terms $I_{1}$,
$I_{1}^{*}$, $I_{2}$ and $I_{2}^{*}$ separately. For $t\gg t_{0}$

\begin{equation}
I_{1}(s)\simeq(\tau_{0})^{\alpha}\frac{2a}{\sqrt{\pi D}}\int_{0}^{\infty}\tau^{-1/2-\alpha}e^{-s\tau}d\tau=(\tau_{0})^{\alpha}\frac{2a}{\sqrt{\pi D}}\frac{\Gamma\left(\frac{1}{2}-\alpha\right)}{s^{1/2-\alpha}}\label{eq:I11}
\end{equation}
which holds for $0<\alpha<1/2.$ Alternatively, if we consider the
limit $s\rightarrow0$ in the exponential term of $I_{1}(s)$, $e^{-s\tau}\simeq1$
and we have

\begin{equation}
I_{1}(s)\simeq(\tau_{0})^{\alpha}\frac{2a}{\sqrt{\pi D}}\int_{\alpha^{-1/\alpha}\tau_{0}}^{\infty}\tau^{-1/2-\alpha}d\tau=\frac{2a\tau_{0}^{1/2}\alpha^{1-\frac{1}{2\alpha}}}{\left(\alpha-1/2\right)\sqrt{\pi D}}\label{eq:I12}
\end{equation}
which holds for $1/2<\alpha<1.$ On the other hand for $t\gg t_{0}$
\begin{equation}
I_{1}^{*}(s)\simeq(\tau_{0})^{\alpha}\frac{2a}{\alpha\sqrt{\pi D}}\int_{0}^{\infty}\tau^{1/2-\alpha}e^{-s\tau}d\tau=(\tau_{0})^{\alpha}\frac{2a}{\alpha\sqrt{\pi D}}\frac{\Gamma\left(\frac{3}{2}-\alpha\right)}{s^{3/2-\alpha}}\label{eq:I1a}
\end{equation}
which holds for $0<\alpha<1.$ With the quantities $I_{1}$ and $I_{1}^{*}$
we can compute the mean occupation time from (\ref{eq:t1}). In particular,
for $0<\alpha<1/2$ and using Eq. (87), (\ref{eq:I11}) and (\ref{eq:I1a})
\[
\left\langle \tilde{T_{1}}(s)\right\rangle \equiv\frac{I_{1}^{*}(s)}{1-\hat{\psi}(s)}\simeq\frac{2a}{\sqrt{\pi D}}\frac{\Gamma\left(\frac{3}{2}-\alpha\right)}{\Gamma\left(1-\alpha\right)s^{3/2}}
\]
and
\[
\left\langle \tilde{T_{2}}(s)\right\rangle \equiv\frac{I_{1}(s)}{s\left(1-\hat{\psi}(s)\right)}\simeq\frac{2a}{\sqrt{\pi D}}\frac{\alpha\Gamma\left(\frac{1}{2}-\alpha\right)}{\Gamma\left(1-\alpha\right)s^{3/2}}.
\]
Adding both terms we readily find
\[
\left\langle T(t)\right\rangle \simeq\frac{2a}{\pi\sqrt{D}}\frac{\Gamma\left(\frac{1}{2}-\alpha\right)}{\Gamma\left(1-\alpha\right)}t^{1/2}\quad\textrm{for}\quad0<\alpha<1/2.
\]
For $1/2<\alpha<1$ we make use of (\ref{eq:I12}) and (\ref{eq:I1a})
to get the same result for $\left\langle \tilde{T_{1}}(s)\right\rangle $
as above but now
\[
\left\langle \tilde{T_{2}}(s)\right\rangle \simeq\frac{2a\tau_{0}^{1/2-\alpha}\alpha^{2-\frac{1}{2\alpha}}}{\sqrt{\pi D}\left(\alpha-1/2\right)\Gamma\left(1-\alpha\right)s^{1+\alpha}}
\]
so that
\[
\left\langle T(t)\right\rangle \simeq\frac{2a\tau_{0}^{1/2}\alpha^{1-\frac{1}{2\alpha}}}{\sqrt{\pi D}\left(\alpha-1/2\right)\Gamma\left(1-\alpha\right)\Gamma\left(\alpha\right)}\left(\frac{t}{\tau_{0}}\right)^{\alpha}\quad\textrm{for}\quad1/2<\alpha<1.
\]

\subsubsection{Mean square occupation time}

We need to compute $I_{2}$ and $I_{2}^{*}$ analogously. First of
all we note from (70) that
\[
\left\langle T^{2}(t)\right\rangle _{\textrm{BM}}\simeq\frac{2a^{2}}{D}t.
\]
If we consider $e^{-s\tau}\simeq1$ in the limit $s\rightarrow0$
the integrals in $I_{2}$ and $I_{2}^{*}$ converge for $\alpha>1$
and $\alpha>2,$ respectively. Then, this approximation does not hold
in our range of interest of the values of $\alpha.$ Instead, we consider
the limit $t\gg t_{0}$ . From (\ref{eq:I2d}) and (\ref{eq:psi})
\begin{equation}
I_{2}(s)\simeq\frac{2a^{2}}{D}(\tau_{0})^{\alpha}\int_{0}^{\infty}e^{-s\tau}\tau^{-\alpha}d\tau=\frac{2a^{2}(\tau_{0})^{\alpha}}{D}\frac{\Gamma\left(1-\alpha\right)}{s^{1-\alpha}}\label{eq:I2}
\end{equation}
which holds for $0<\alpha<1.$ Analogously, from (\ref{eq:I2ad})
and (\ref{eq:S})
\begin{equation}
I_{2}^{*}(s)\simeq\frac{2a^{2}(\tau_{0})^{\alpha}}{\alpha D}\int_{0}^{\infty}e^{-s\tau}\tau^{1-\alpha}d\tau=\frac{2a^{2}(\tau_{0})^{\alpha}}{\alpha D}\frac{\Gamma\left(2-\alpha\right)}{s^{2-\alpha}}\label{eq:I2a}
\end{equation}
which holds also for $0<\alpha<1.$ The first and third terms of (\ref{eq:t22})
are of the same order and both behave as $s^{-2}$ in the limit $s\rightarrow0$.
These terms can be added using (\ref{eq:I2}) and (\ref{eq:I2a})
to find
\begin{equation}
\frac{I_{2}^{*}(s)}{1-\hat{\psi}(s)}+\frac{I_{2}(s)}{s\left(1-\hat{\psi}(s)\right)}=\frac{2a^{2}}{Ds^{2}}.\label{eq:i}
\end{equation}
Plugging (\ref{eq:I11}), (\ref{eq:I12}), (\ref{eq:I1a}) together
with (\ref{eq:i}) into the expression (\ref{eq:t22}), we find
\begin{equation}
\left\langle T^{2}(t)\right\rangle \simeq\left\{ \begin{array}{cc}
\frac{2a^{2}}{D}\left[1+\frac{2\alpha\Gamma^{2}\left(\frac{1}{2}-\alpha\right)}{\pi\Gamma^{2}\left(1-\alpha\right)}\right]t, & 0<\alpha<1/2\\
\frac{8a^{2}\tau_{0}}{\pi D}\frac{\alpha^{4-\frac{1}{\alpha}}}{\left(\alpha-1/2\right)^{2}\Gamma^{2}\left(1-\alpha\right)\Gamma(1+2\alpha)}\left(\frac{t}{\tau_{0}}\right)^{2\alpha}, & 1/2<\alpha<1
\end{array}\right..\label{eq:t2f}
\end{equation}

\subsubsection{EB}

From the definition of the ergodicity breaking parameter EB given
in Eq. (62) one has
\[
\textrm{EB}=\frac{\left\langle T^{2}(t)\right\rangle }{\left\langle T(t)\right\rangle ^{2}}-1.
\]
In the long time limit we can make use of the expressions above to
find EB:

\[
\textrm{EB}=\left\{ \begin{array}{cc}
\alpha\pi-1+\frac{1}{2}\left[\frac{\pi\Gamma\left(1-\alpha\right)}{\Gamma\left(\frac{1}{2}-\alpha\right)}\right]^{2}, & 0<\alpha<1/2\\
\frac{2\Gamma^{2}(1+\alpha)}{\Gamma(1+2\alpha)}-1, & 1/2<\alpha<1
\end{array}\right.
\]
This is reported in the main text in Eq. (\ref{eqtttt1}).

\end{widetext}

\end{document}